\documentclass[preprint,amssymb, amsmath,aps,onecolumn]{revtex4-1}
\usepackage{graphicx}
\usepackage{subfigure}
\usepackage{natbib}
\usepackage{amsmath}
\usepackage{epsfig}
\usepackage{docs}
\usepackage{bm}
\usepackage{color}
\usepackage{transparent}
\usepackage[colorlinks=true,linkcolor=blue]{hyperref}
\expandafter\ifx\csname package@font\endcsname\relax\else
 \expandafter\expandafter
 \expandafter\usepackage
 \expandafter\expandafter
 \expandafter{\csname package@font\endcsname}
\fi
\usepackage{wrapfig}
\usepackage{rotating}

\begin{document}

\title[]{
Influence of a chemical reaction on viscous fingering: \\
Effect of the injection flow rate}

\author{Priyanka Shukla$^{1}$\footnote{priyanka@iitm.ac.in} and A. De Wit$^2$\footnote{adewit@ulb.ac.be}}
\affiliation{$^1$Department of Mathematics,\\
 Indian Institute of Technology Madras, Chennai - 600036, India \\
$^2$Universit{\' e} libre de Bruxelles, Nonlinear Physical Chemistry Unit, CP 231, Facult{\' e} des Sciences, 
Campus Plaine, 1050 Brussels, Belgium
 }

\date{\today}

\begin{abstract}
The hydrodynamic viscous fingering instability can be influenced by a simple viscosity changing chemical reaction of type $A+B \rightarrow C$, when a solution of reactant $A$ is injected into a solution of $B$ and a product $C$ of different viscosity is formed. We investigate here numerically such reactive viscous fingering in the case of a reaction decreasing the viscosity to define the optimal conditions on the chemical and hydrodynamic parameters for controlling fingering. In particular, we analyze the influence of the injection flow rate or equivalently of the P{\' e}clet number (${\rm Pe}$) of the problem on the efficiency of the chemical control of fingering. We show that the viscosity decreasing reaction has an increased stabilizing effect when ${\rm Pe}$ is decreased. On the contrary, fingering is more intense and the system more unstable when ${\rm Pe}$ is increased. The related reactive fingering patterns cover then respectively a smaller (larger) area than in the non-reactive equivalent. Depending on the value of the flow rate, a given chemical reaction may thus either enhance or suppress a fingering instability. This stabilization and destabilization at low and high ${\rm Pe}$ are shown to be related to the ${\rm Pe}$-dependent characteristics of a minimum in the viscosity profile that develops around the miscible interface thanks to the effect of the chemical reaction.
\end{abstract}

\maketitle


\section{Introduction}
\label{sec:intro}

A hydrodynamic viscous fingering (VF) instability can deform the interface between two different fluids when a high mobility fluid of lower viscosity displaces a more viscous and hence less mobile one in a porous medium \cite{ST1958,S1986,TH1986,Homsy1987,TH1988}.  In numerous industrial and environmental problems such as enhanced oil recovery, ${\rm CO}_2$ sequestration, combustion, hydrology, soil remediation, etc.~\cite{OT1984,LK1999,FCJ2013,Farajzadeh2015,RDNS1998}, this fingering instability can interplay with chemical reactions. In the past few decades, viscous fingering has been analyzed in reactive systems on both miscible and immiscible interfaces~\citep{HB1995,JH2000,NU2001,NU2003,NOKT2009,WH1999, WH1999a, FH2003,PSZB2007,NMKT2007, GW2009, NKKT2009,HTAW2010, HA2010a,HA2010b,NW2011,NKKT2011,RNIMTWit2012,riolfo,alh13,Nagatsu2010}. If the reaction does not modifies the viscosity {\it in-situ}, the chemical species are passively advected by the flow and the fingering properties of the interface remain similar to those of the nonreactive system~\cite{HB1995,JH2000, NU2001,NU2003,NOKT2009}. The flow in the fingering patterns can on the other hand change the spatio-temporal distribution of the reactants and influence the yield of the reaction. Active influence of chemistry on fingering can be obtained as soon as the chemical reaction taking place around the interface between the two fluids modifies their physical properties and, in particular, their viscosity~\cite{dew16}. The reaction then influences the stability as well as the spatio-temporal dynamics of the flow.  In turn, the hydrodynamic flow affects mixing and thus the amount and spatial distribution of chemical species and a highly nonlinear feedback is established between chemistry and hydrodynamics.

For cases where reactions actively change the viscosity {\it in-situ}, numerical simulations have first shown on the basis of a bistable chemical reaction scheme that the properties of miscible VF are modified when the reaction changes the viscosity across the reactive miscible interface~\cite{WH1999,WH1999a}. The bistable nature of chemical kinetics is then responsible for a new phenomenon of droplet formation isolating regions of high or low viscosity within connected domains of the other steady state. In other studies, the active influence of $A+B \rightarrow C$ types of chemical reaction on miscible viscous fingering has been studied both experimentally~\cite{PSZB2007,NMKT2007,NKKT2009,NKKT2011,RNIMTWit2012,riolfo} and theoretically~\cite{GW2009,HTAW2010,HA2010a,HA2010b,NW2011,RNIMTWit2012, alh13}. \citet{PSZB2007} have in particular studied experimentally chemically-driven fingering at the miscible reactive interface between two aqueous solutions of same viscosity when a reaction between a cationic surfactant and an organic salt produces an elastic more viscous worm-like micellar fluid. Various fingering regimes have been identified depending on concentrations, fluid characteristics and injection flow rate (or equivalently  P{\' e}clet number, defined as the ratio of the convective to diffusive transport rates).  

In some experiments by Nagatsu et al., a less-viscous acidic or basic aqueous solution was injected into a more-viscous polymeric solution, the viscosity of which depends on pH ~\cite{NMKT2007,NKKT2009,NKKT2011}. It is observed that, when the viscosity is increased (decreased) by the reaction, fingers are widened (narrowed), which is mainly due to suppressed (enhanced) shielding effects. Interestingly, opposite results have been observed at moderate reaction rates for systems with a viscosity decrease~\cite{NKKT2009} and increase~\cite{NKKT2011}. In the case where the non reactive displacement is stable (more viscous solution displacing a less viscous once), it has even been shown experimentally that the reaction is able to trigger VF~\cite{RNIMTWit2012}. Depending whether the reaction increases or decreases viscosity, a different fingering pattern is then obtained.

The experimental study of Nagatsu et al.~\cite{NMKT2007} showed that at `large' injection rate, or equivalently high P{\' e}clet number (${\rm Pe}$), an instantaneous chemical reaction can have opposite effects on miscible VF when a less viscous (acidic or basic) solution is injected radially into a more viscous (e.g.~polymeric solution) one in a Hele-Shaw cell depending whether the reaction locally increases or decreases the viscosity. In the viscosity increase case, the VF pattern is ``denser" in the sense that it covers a more compact area in the Hele-Shaw cell than the non-reactive pattern. On the contrary, a VF pattern covering a smaller area (also qualified as ``less dense pattern") was reported in the viscosity decrease reactive case. Recently, new experiments have been carried out focusing on the influence of the injection rate on viscosity increasing and decreasing reactive systems~\cite{riolfo,NMW2015}. Interestingly, it was found that, at lower Pe, the trends are opposite than at high ${\rm Pe}$ i.e. for viscosity decreasing reactions, the system can be  stabilized at low injection flow rates.  These experiments~\cite{NMKT2007,riolfo,NMW2015} thus clearly show that, in the presence of a viscosity decreasing reaction, the reactive VF patterns can be controlled by varying the P{\' e}clet number. Moreover, when the reaction induced viscosity decrease is large enough, a suppression of the VF instability can be obtained at small ${\rm Pe}$. In numerical studies, the explicit influence of the injection rate on reactive VF has however not been addressed explicitly. 

In this context, our objective is here to analyze numerically the influence on the VF instability of changes in the injection flow rate i.e. changes in the  ${\rm Pe}$ number of the problem when a simple $A+B \rightarrow C$  chemical reaction  decreases the viscosity {\it in situ}. To this end, we integrate numerically the  reaction-diffusion-convection (RDC) equations of reactive VF in porous media and analyze the properties of the  fingering patterns for different values of ${\rm Pe}$. We show that a viscosity-decreasing reaction enhances stabilization or destabilization of the interface at respectively low and high ${\rm Pe}$, with regard to the non-reactive system. This is related to the possibility at low ${\rm Pe}$ for chemistry to build up a minimum in the viscosity profile that blocks the further progression of fingering and stabilizes the system. On the contrary, at high ${\rm Pe}$, chemistry does not have time to act to decrease the viscosity and the classical enhanced destabilization when the flow rate is increased is then observed. 

These results highlights the optimum conditions on flow conditions to obtain stabilization by reactions of VF. This is of practical importance as it paves the way to a possible  chemical control of fingering instabilities appearing in many practical situations ranging from geophysical to environmental problems. 

This paper is organized as follows. The problem description and the related RDC model are given in Sec.~\ref{sec:probdes_eqns}. In Sec.~\ref{subsec:numerical-method}, the numerical method used to integrate the model is discussed. The characteristics of VF patterns and in particular the influence of the P{\' e}clet number are studied in Sec.~\ref{sec:results}. The non-reactive and reactive cases are given in \ref{subsec:conc_NR} and \ref{subsec:VF_reactive}, respectively. A quantitative analysis and parametric study are carried out in Secs.~\ref{subsec:quantitative} and~\ref{sec:parametric_analysis}. At the end, conclusions and outlook are given in Sec.~\ref{sec:Conclusion and Outlook}.

\section{Problem description and governing equations}
\label{sec:probdes_eqns}

Consider a homogeneous two-dimensional porous medium  or horizontal thin Hele-Shaw cell of length $L_x$ and width $L_y$ with constant permeability $\kappa$ in which a miscible solution of reactant $A$ with viscosity $\mu_A$ is injected from left to right  into a solution of reactant $B$ with viscosity $\mu_B$ at a constant speed $U$ along the $x$-direction (Fig.1). We assume that the initial concentrations of $A$ and $B$ are both equal to $a_0$. The initial position of the miscible interface is $x_0$. 
\begin{figure}[!ht]
\begin{center}
\includegraphics[scale=0.5]{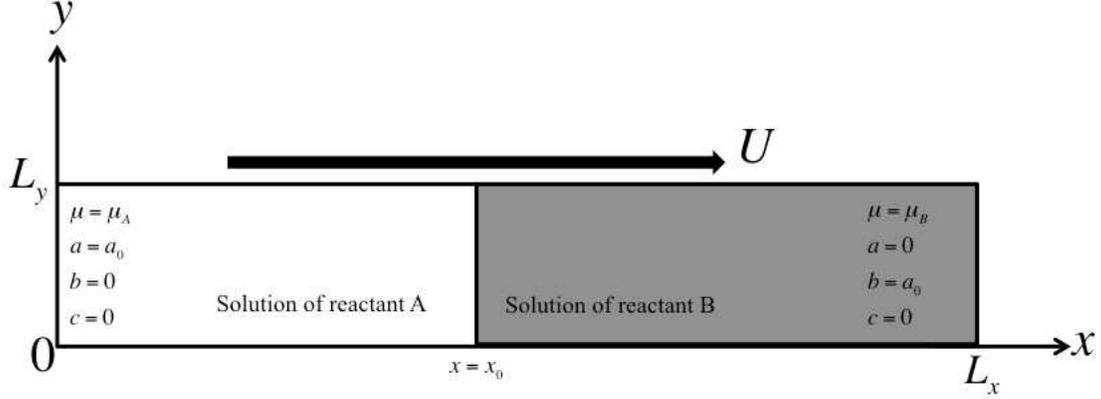}\\
\caption{\small{Sketch of a two-dimensional porous medium of length $L_x$ and width $L_y$ with permeability $\kappa$ in which a solution of reactant $A$ with viscosity $\mu_A$ is displacing a solution of reactant $B$ of viscosity $\mu_B$ from left to right at a constant speed $U$. Here $x_0$ and $a_0$ are the initial contact position and initial concentration of reactants, respectively.
}}
\label{fig:sketch}
\end{center}
\end{figure}
Upon contact between the two solutions, a simple $A+B \rightarrow  C$ chemical reaction takes place in the miscible interface zone where $A$ and $B$ meet by diffusion, react, and yield the product $C$ of viscosity $\mu_C$. The objective is to analyze numerically how the dynamic decrease of viscosity driven by the reaction can influence the VF instability and in particular what is the influence of the injection speed $U$ on this effect.

To analyze the problem, the system is considered as incompressible and neutrally buoyant. The dynamics is modeled using Darcy's law  for the velocity field along with three reaction-diffusion-convection (RDC) equations for the concentrations: 
\begin{align}
\boldsymbol{\nabla}\cdot\boldsymbol{u}&=0,
\label{eqn:Mass}
\\
\boldsymbol{\nabla}p&=-\frac{\mu(a,b,c)}{\kappa}\boldsymbol{u},
\label{eqn:Momentum}
\\
\frac{\partial a}{\partial t}+\boldsymbol{u}\cdot\boldsymbol{\nabla}a &= D_{\!A}\,\nabla^2 a-k\,a\,b,
\label{eqn:concA}
\\
\frac{\partial b}{\partial t}+\boldsymbol{u}\cdot\boldsymbol{\nabla}b &= D_{\!B}\,\nabla^2 b-k\,a\,b,
\label{eqn:concB}
\\
\frac{\partial c}{\partial t}+\boldsymbol{u}\cdot\boldsymbol{\nabla}c &= D_{\!C}\,\nabla^2 c+k\,a\,b,
\label{eqn:concC}
\end{align}
where $a$, $b$, and $c$ denote the concentrations of the reactants $A$ and $B$ and of the product $C$, respectively, $k$ is the kinetic constant, $p$ is the pressure, $D_{A}$, $D_B$ and $D_C$ are the diffusivities of the reactants $A$ and $B$ and the product $C$, respectively, $\boldsymbol{u}=(u,v)$ is the two-dimensional flow velocity and $\kappa$ is the constant permeability. The viscosities of the solution when only one species is present in concentration $a_0$ are defined as $\mu_A$, $\mu_B$ and $\mu_C$, respectively in the presence of the reactants $A$, $B$ or of the product $C$. Following previous theoretical work on viscous fingering~\citep{TH1986,WH1999a,WH1999,MMW2007,GW2009,HTAW2010,
HA2010a,HA2010b,NW2011,riolfo,alh13}, we assume the viscosity as an exponential function of the concentrations of $A$, $B$ and $C$ as
\begin{equation}
\mu(a,b,c) = \mu_A \,e^{[R_b b + R_c c]/a_0},
\label{eqn:viscosity}
\end{equation}
where $R_b$ and $R_c$ are the log-mobility ratios defined as
\begin{equation}
R_b= \mbox{ln} \left( \frac{\mu_B}{\mu_A}\right)\qquad \mbox{and}\qquad  R_c= \mbox{ln} \left( \frac{\mu_C}{\mu_A}\right).
\end{equation}
For the non-reactive VF case or the equivalent specific reactive case when the product $C$ has the same viscosity as one of the reactant (i.e.~$R_b =R_c$), the system is  unstable when the lower viscosity solution of A displaces the more viscous solution of B i.e. when $\mu_A<\mu_B$ or $R_b>0$. Let us  analyze how this stability is changed when both $\mu_C$ and the injection speed $U$ are varied.

\subsection{Non-dimensional Equations}

To specifically let the injection speed appear in the dimensionless problem under the form of a P\'eclet number, the reference scales for length, velocity, time, concentration, viscosity, diffusivity and pressure are taken as $L_y$, $U$, $L_y/U$, $a_0$, $\mu_A$, $D_C$ and $\mu_AU L_y/\kappa$, respectively. For simplicity, equations are written in a reference frame moving with speed $U$ by transforming variables as $\boldsymbol{x} \rightarrow \boldsymbol{x} - U t  \boldsymbol{e_x}$ and $\boldsymbol{u} \rightarrow \boldsymbol{u} - U \boldsymbol{e_x}$ with $\boldsymbol{e_x}$ being the unit vector along $x$ direction. The dimensionless form of (\ref{eqn:Mass})--(\ref{eqn:viscosity}) can then be written as
\begin{align}
\boldsymbol{\nabla}\cdot \boldsymbol{u} &=0,
\label{eqn:mass_nd}\\
\boldsymbol{\nabla} p &= -\mu(a,b,c) (\boldsymbol{u}+\boldsymbol{e_x}),
\label{eqn:momentum_nd}\\
\frac{\partial a}{\partial t}+ \boldsymbol{u} \cdot \boldsymbol{\nabla} a &= \delta_a {\rm Pe}^{-1}  \nabla^2 a -D_a\,a\,b,
\label{eqn:a_nd}\\
\frac{\partial b}{\partial t}+ \boldsymbol{u} \cdot \boldsymbol{\nabla} b &=  \delta_b{\rm Pe}^{-1} \nabla^2 b -D_a\,a\,b,
\label{eqn:b_nd}\\
\frac{\partial c}{\partial t}+ \boldsymbol{u} \cdot \boldsymbol{\nabla} c &=  {\rm Pe}^{-1}  \nabla^2 c +D_a\,a\,b,
\label{eqn:c_nd}\\
\mu(a,b,c) &=e^{(R_b b + R_c c)},
\label{eqn:kappa_nd}
\end{align}
where $D_a\!=\!k a_0 L_y/U \!=\! \tau_h/\tau_c$ is the dimensionless Damk{\"o}hler number 
defined as the ratio of the hydrodynamic time scale $\tau_h\!=\!L_y/U$ to the chemical 
time scale $\tau_c\!=\!1/k a_0$. The P\'eclet number ${\rm Pe}\!=\!U L_y/D_c=\tau_h/\tau_D$ is the ratio of the convective time $\tau_h$ to the diffusive time $\tau_D=D_c/U^2$ while $\delta_a\!=\!D_A/D_C$ and $\delta_b\!=\!D_B/D_C$ are the diffusion coefficient ratios. Taking the curl of the momentum equation and defining the stream function $\psi(x,y)$ as $u \!=\! \partial \psi/\partial y$ and $v\!=\!- \partial \psi/\partial x$, we get
 \begin{eqnarray}
\nabla^2  \psi&=&  R_b ( \psi_x b_x + \psi_y b_y + b_y)  +R_c ( \psi_x c_x + \psi_y c_y + c_y),
\label{eqn:mass}\\
a_t+  a_x \psi_y - a_y \psi_x  &=&   \delta_a{\rm Pe}^{-1} \nabla^2 a -D_a\,a\,b,
\label{eqn:a1}\\
b_t+  b_x \psi_y - b_y \psi_x  &=& \delta_b{\rm Pe}^{-1}  \nabla^2 b -D_a\,a\,b,
\label{eqn:b1}\\
c_t+ c_x \psi_y - c_y \psi_x  &=& {\rm Pe}^{-1}  \nabla^2 c +D_a\,a\,b,
\label{eqn:c1}
\end{eqnarray}
where the subscripts $x$ and $t$ represent the respective derivatives. The last term in \eqref{eqn:a1}--\eqref{eqn:c1} corresponds to the reaction rate $\mathcal{R}$:
\begin{equation}
\mathcal{R}(x,y,t) = D_a\,a(x,y,t)\,b(x,y,t).
\label{eqn:reactionrate}
\end{equation}
Comparing the present RDC model \eqref{eqn:mass}--\eqref{eqn:c1} with those previously studied in the literature~\cite{TH1988,GW2009,HTAW2010,NW2011,PM2015},  
we note that: (i) when $D_a\!=\!0$ we recover the classical model for non-reactive viscous fingering similar to the one studied by Tan and Homsy \cite{TH1988,PM2015}; 
 (ii) when $D_a\!\neq\!0$, $Pe\!=\!1$ and $R_b\!=\!0$ we obtain the model of reactive VF for solutions of A and B of same viscosity as analyzed numerically by~\citet{GW2009}; (iii) when $D_a\!\neq\!0$, ${\rm Pe}=\delta_a\!=\!\delta_c\!=\!1$ we get back to the reactive VF model with $A$, $B$ and $C$ of different viscosity but species diffusing all at the same rate as studied by~\citet{HTAW2010} and~\citet{NW2011}.

As the dynamics of the reactive zone is independent of boundary conditions as long as the unstable fingered front does not confront its periodic extension~\cite{TH1988}, 
we use periodic boundary conditions in both  directions. 
The initial conditions for the stream function and product concentration $c$ are taken as $\psi(x,y)\!=\!0$ and $c(x,y)\!=\!0$, for all $(x,y)$, respectively. For the initial concentrations of the reactant $A$ and $B$ solutions, we use a step front between  $A=1, B=0$ on the left and $B=1, A=0$ on the right of $x\!=\!x_0$ with a random noise of amplitude of order $10^{-2}$ added in the front to trigger the instability. The dimensionless system size is $\mathcal{A} \times 1$, where $\mathcal{A}=L_x/L_y$ is the aspect ratio. 
Equations~\eqref{eqn:mass}--\eqref{eqn:c1} together with the initial and boundary conditions form an initial-boundary value problem with six dimensionless control parameters---namely, $R_b$, $R_c$, $D_a$, $\delta_a$, $\delta_b$ and ${\rm Pe}$. To decrease the wide range of possibilities, we fix here  $\delta_a=\delta_b=1$ to focus on the effect of the reaction (variable $Da$ and $R_c$ for a given $R_b$) and flow speed (variable ${\rm Pe}$) on the fingering instability.

\subsection{Numerical Method}
\label{subsec:numerical-method}

To solve \eqref{eqn:mass}--\eqref{eqn:c1}, we use a pseudo-spectral numerical scheme  based on the discrete Fourier transform library FFTW 3.3.4~\citep{TH1988,Fornberg1998,WH1999,WH1999a,GW2009}. In order to avoid any interaction between the unstable fingered front and its periodic extension, we choose a domain with a large aspect ratio. The physical and computational domain size ($L_x\times L_y$) are $32 \times1$ and $4098\times 128$, respectively. The time step of numerical integration is chosen as ${\rm dt} = 10^{-4}$. To validate our code, we have successfully reproduced previous  nonlinear simulation results of 
non-reactive~\citep{TH1988,PM2015} and  reactive~\citep{GW2009,HTAW2010,NW2011} systems.

\section{Results}
\label{sec:results}



\subsection{Non-reactive system}
\label{subsec:conc_NR}

\begin{figure}[!htbp]
\begin{center}
\includegraphics[scale=0.5]{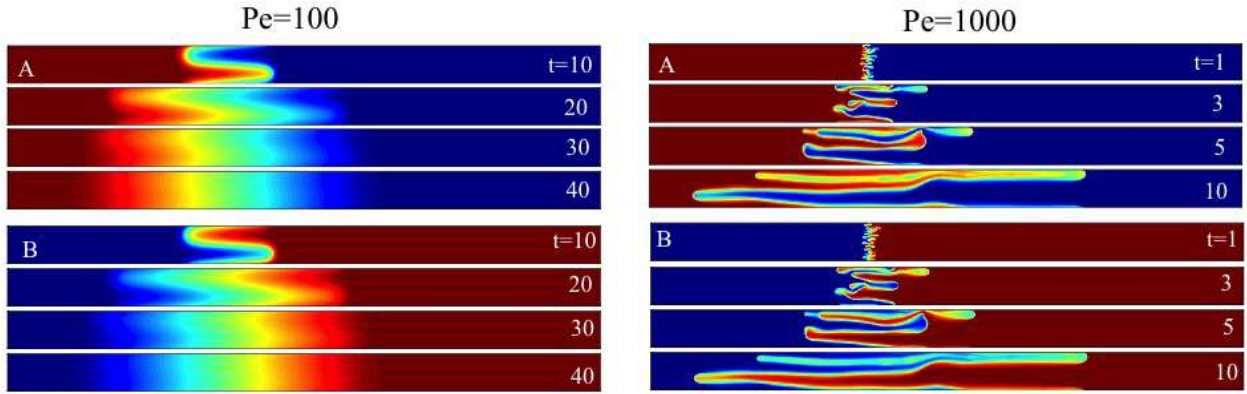}
\caption{\small{
Equivalent non-reactive ($R_b=R_c=2$) system : 
Concentrations of $A$ and $B$ for ${\rm Pe}=100$ [left] and $1000$ [right] at four different times (from top to bottom). Concentration fields are scaled between zero (blue) and one (red). The viscosity (not shown) varies in a similar way as the concentration of $B$.}}
\label{fig:NR_pe100_1000}
\end{center}
\end{figure}

It is already  known that, in absence of any reaction effect ($Da=0$ or $R_b=R_c$ \cite{HTAW2010,NW2011}), increasing the injection speed (i.e. increasing ${\rm Pe}$ in our dimensionless formulation of the problem) increases the destabilization of the interface by VF when $R_b > 0$ \citep{TH1988,Homsy1987,PM2015}. As a reference case, this observation is shown in Fig.~\ref{fig:NR_pe100_1000} which illustrates the concentration of reactants ($A$ and $B$) for ${\rm Pe}=100$ and ${\rm Pe}=1000$, respectively, at four different times. As ${\rm Pe}$ increases, fingering becomes more intense and the wavelength of the pattern decreases as the interface becomes more unstable. It is also observed that, at low Pe,  the deformed interface tends to flatten as time evolves thanks to  transverse diffusion.

\begin{figure}[!htbp]
\begin{center}
\includegraphics[scale=0.35]{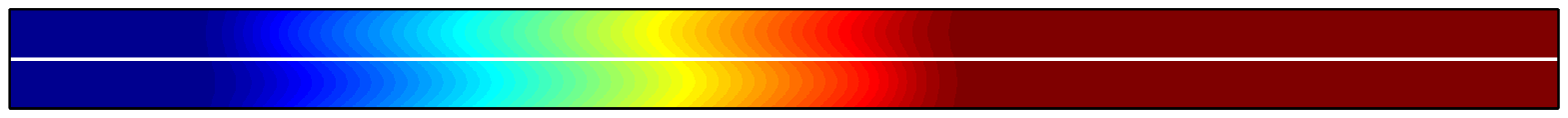}\qquad\quad
\includegraphics[scale=0.35]{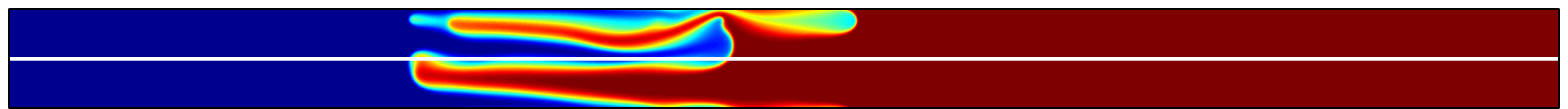}
\includegraphics[scale=0.35]{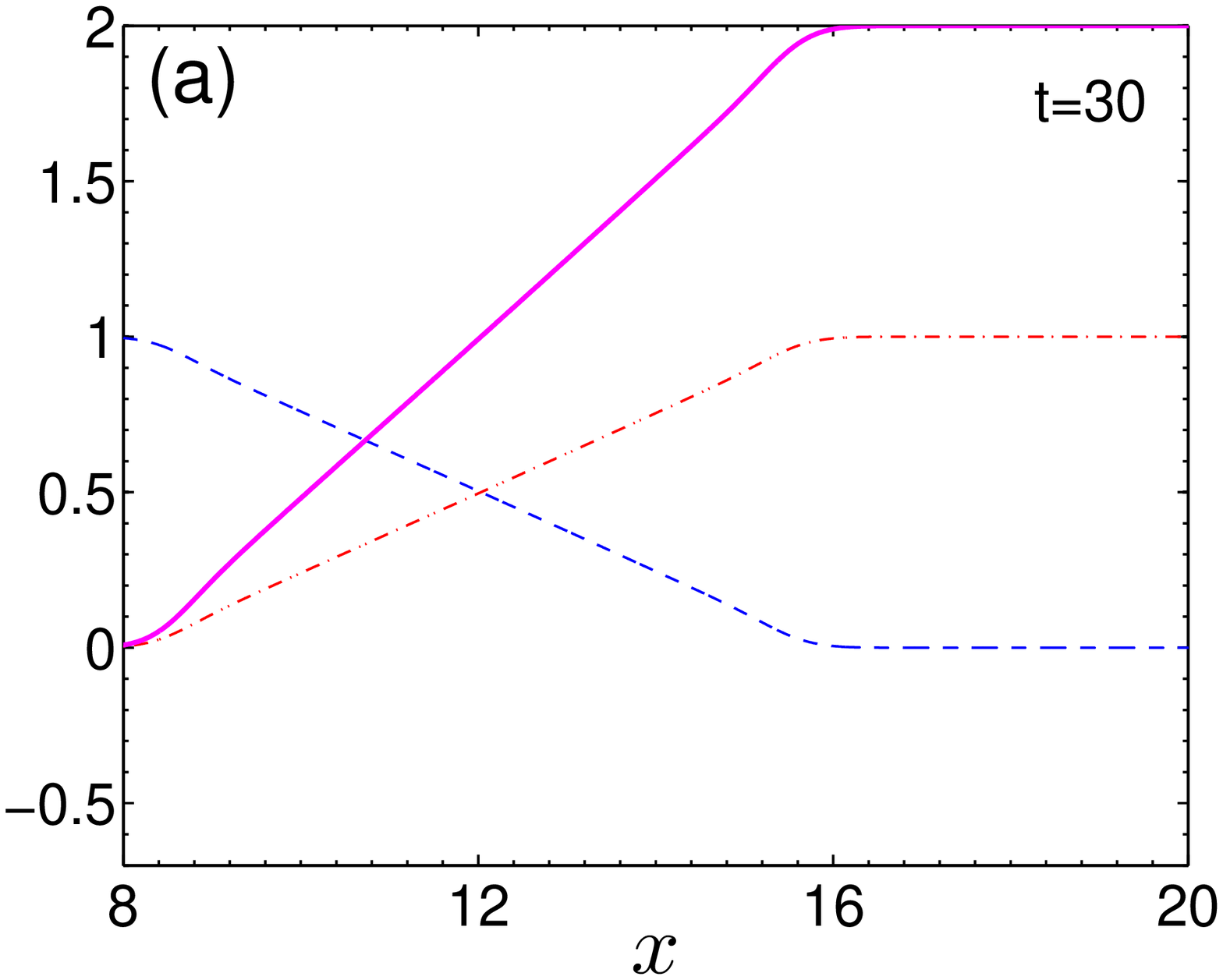}
 \includegraphics[scale=0.35]{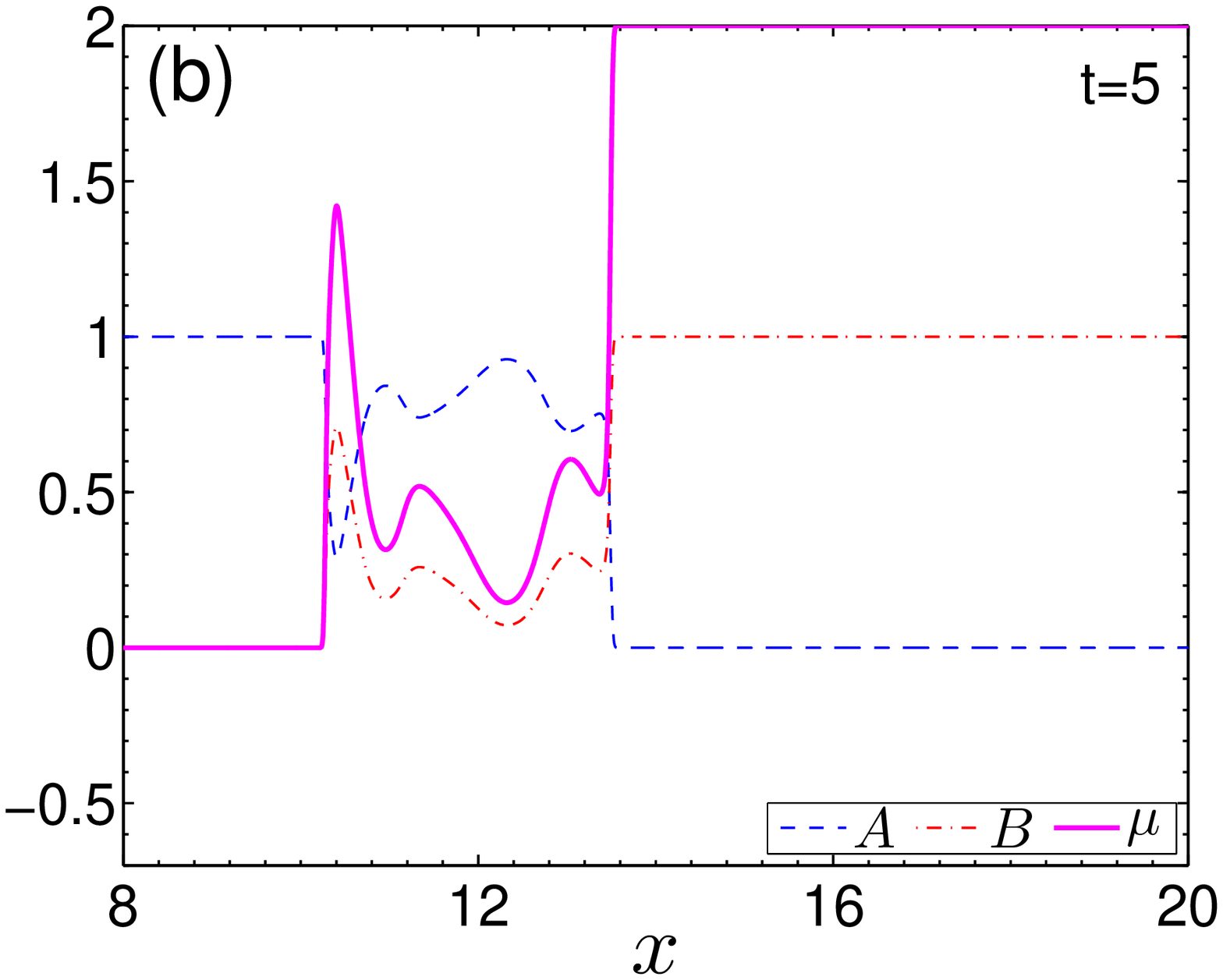}
\caption{\small{Spatial profiles of concentrations of $A$ (dashed blue line), $B$ (dash-dotted red line), and of $\mbox{ln}(\mu)$ (solid magenta line) along the injection direction at $y=L_y/2$ for $R_b=R_c=2$ and (a) ${\rm Pe}=100$ at $t=30$ or (b) ${\rm Pe}=1000$ at $t=5$, see Fig.~\ref{fig:NR_pe100_1000}. The top figures represent the corresponding two-dimensional map of $\mbox{ln}(\mu)$ through which the one dimensional sections are shown.
 }}
\label{fig:Fixed_positionConc_NR}
\end{center}
\end{figure}

Figure~\ref{fig:Fixed_positionConc_NR} compares the one dimensional profiles of the concentrations of $A$ and $B$, and the logarithm of viscosity $\mbox{ln}(\mu)$ at a fixed transverse location $y=L_y/2$ for ${\rm Pe}=100$ at time $t=30$ and ${\rm Pe}=1000$ at $t=5$, respectively. As a reference, the white line $y=L_y/2$ is shown in the corresponding two-dimensional map of $\mbox{ln}(\mu)$ on the top of the panels. While the concentrations and viscosity profiles at large ${\rm Pe}$ show bumps characteristic of the fingering instability, these profiles are quasi linear between the end-point values at small ${\rm Pe}$ indicating a more stable interface. This stabilizing effect at low Pe is in agreement with previous results \cite{TH1986,TH1988,PM2015}.

\subsection{Reactive system}
\label{subsec:VF_reactive}
Let us now analyze the effect of ${\rm Pe}$ on  reactive VF when an $A+B \rightarrow C$ reaction produces the product $C$ of lower viscosity (negative value of $R_c$) such that the viscosity of the system develops in time a minimum around the reactive front.
\begin{figure}[!htbp]
\begin{center}
\includegraphics[scale=0.46]{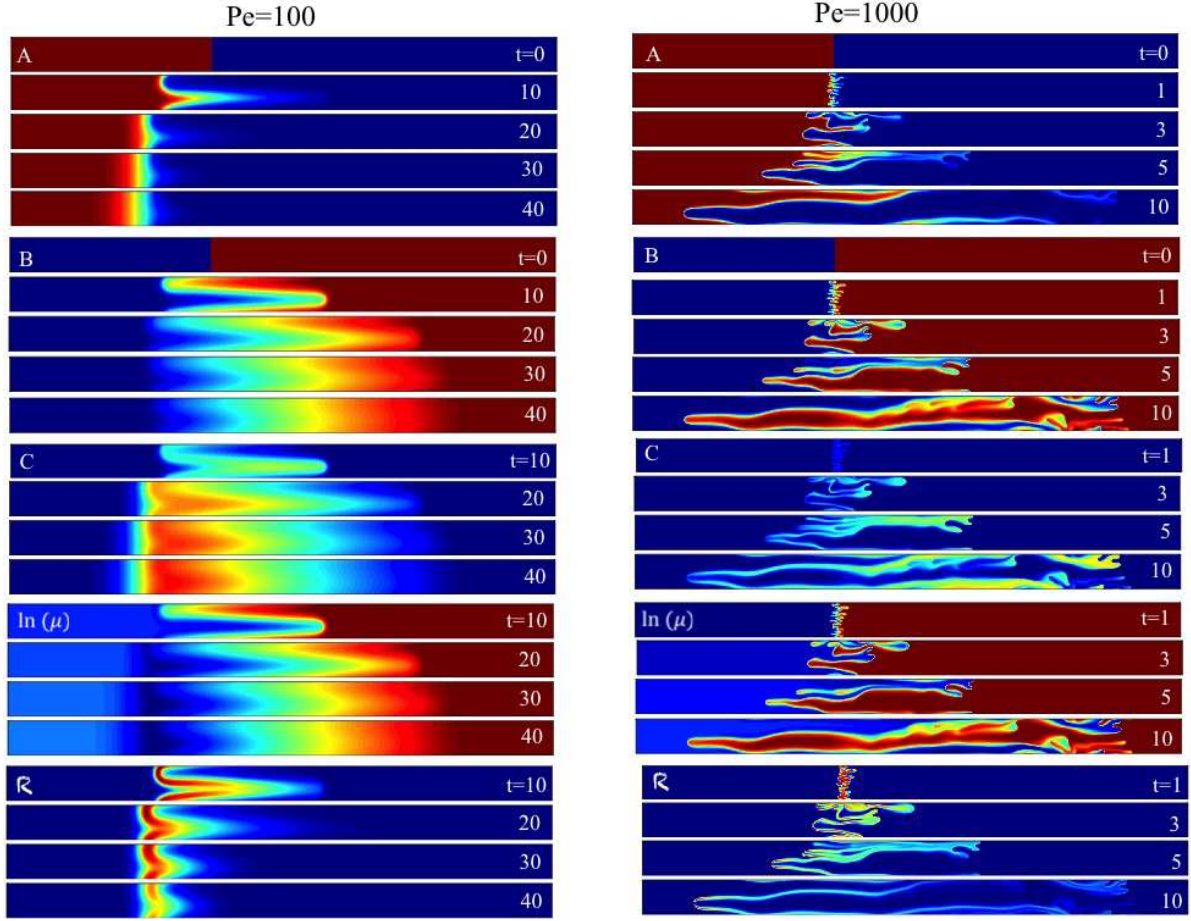}
\caption{\small{Reactive VF at $R_b=2$, $R_c=-2$ and $D_a=1$.  
The first and second columns represent the concentrations of $A$, $B$, $C$, viscosity in log-scale (${\rm ln}(\mu)$) and $\mathcal{R}$ for ${\rm Pe}=100$ and $1000$ at various times, respectively.  
$A$ and $B$ are scaled between zero (blue) and one (red), $C$ is scaled between zero (blue) and 0.5 (red), and ${\rm ln} (\mu)$ and $\mathcal{R}$ are shown in their absolute values. }}
\label{fig:reactive_case_Da1_Pe100_1000}
\end{center}
\end{figure}

Figure~\ref{fig:reactive_case_Da1_Pe100_1000} shows the concentrations, ${\rm ln}(\mu)$, and reaction rate $\mathcal{R}$ at $D_a=1$, $R_b=2$ and $R_c=-2$ for two values of P{\'e}clet numbers ${\rm Pe}=100$ (first column) and $1000$ (second column). Two opposite behaviours are obtained at low and high Pe: at ${\rm Pe}=1000$, fingering is more intense than in the non reactive case with coarsening, and more repetitive shielding and tip splitting \cite{NW2011}. The fingered zone extends on a larger spatial extent than in the non reactive case (Fig.2) suggesting that the reaction has here a destabilizing effect. A comparison of the transverse averaged viscosity profile in the non reactive (Fig.3b) and reactive (Fig.5b) cases shows that, at ${\rm Pe}=1000$, the decrease in viscosity induced by the reaction leads to a sharper viscosity jump which can explain the increased destabilisation. As a consequence, fingering extends both in the $A$- and $B$-rich regions with the reaction rate being localised at the fingered  frontier between the two reactants. On the contrary, at ${\rm Pe}=100$ (Fig.4, first column), a minimum in viscosity develops in the course of time where the less viscous $C$ separates the two reactants $A$ and $B$ (Fig.5a). The reaction rate correspondingly decreases in time and remains strongly localised at a given location. The time scales are also longer as more time is needed to cover the same distance. Interestingly, fingering is weak and  remains longer in the boundary zone where the less viscous $C$ displaces the more viscous $B$ then in the  stable part of the non-monotonic profile where A pushes the less viscous $C$. This means that, in experiments where often a dye is used to visualize the fingering pattern, the instability would quickly become unnoticeable if the dye is diluted in the injected A reactant~\cite{riolfo}. 

A comparison of the spatio-temporal distribution of $A$ in Fig.2 (non reactive) and~4 (reactive) leads thus to the conclusion that, at high ${\rm Pe}$, reactive fingering is more intense with more ramified fingers that cover a larger area  in the presence of reaction. On the contrary, at low 
${\rm Pe}$, fingering is stabilized by the reaction. The effect of the reaction decreasing the viscosity has thus an opposite effect on the flow at high and low Pe, as observed experimentally~\cite{riolfo,NMKT2007}.

\begin{figure}[!htbp]
\begin{center}
\includegraphics[scale=0.44]{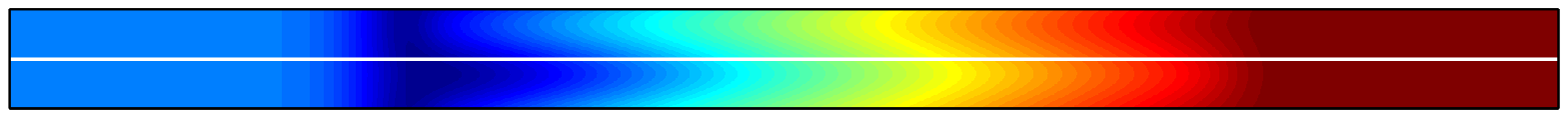}
\includegraphics[scale=0.44]{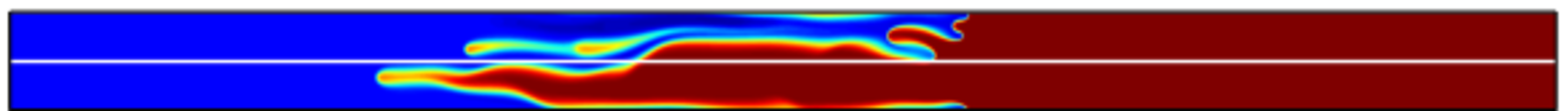}
\includegraphics[scale=0.35]{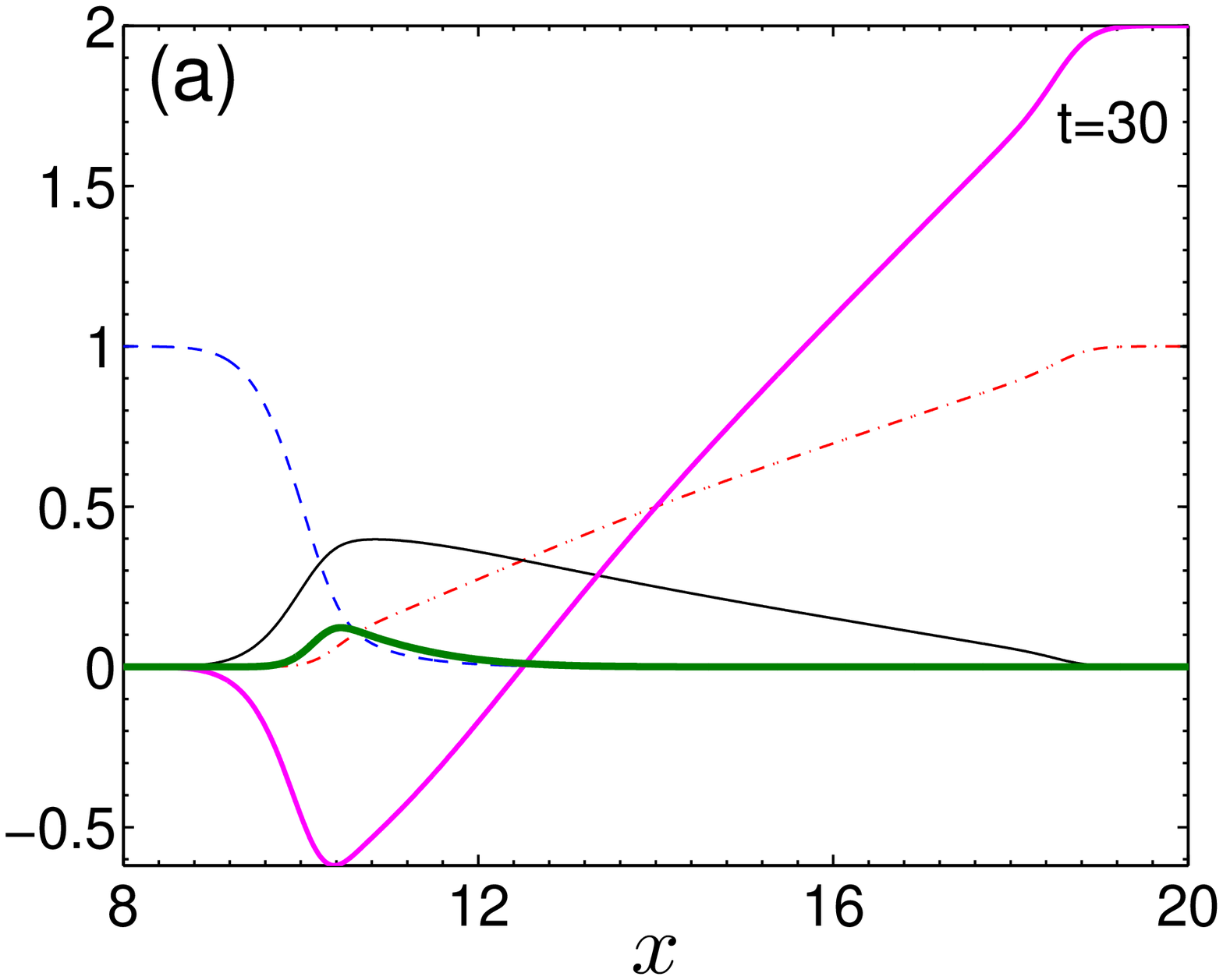}
 \includegraphics[scale=0.35]{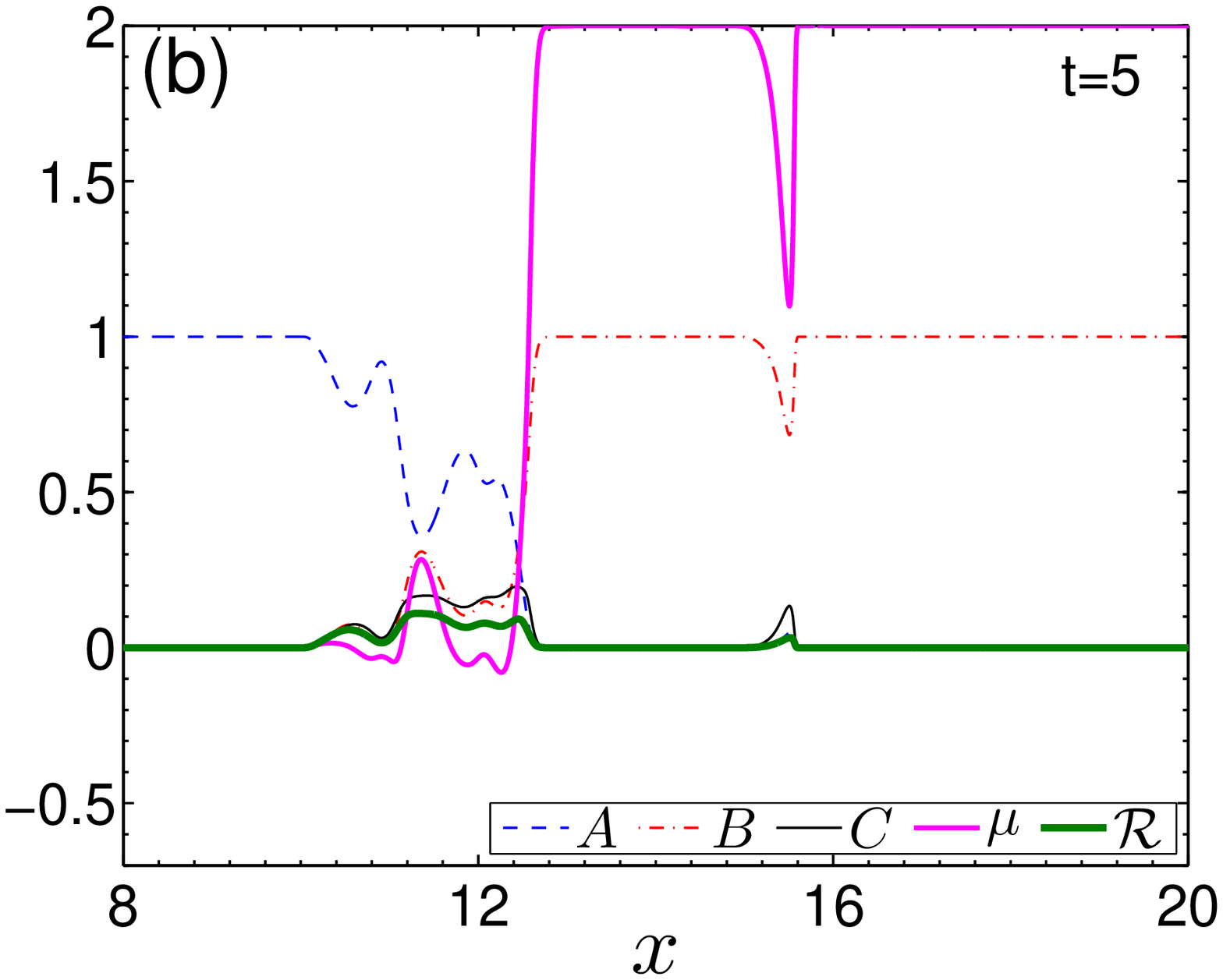}
\caption{\small{
Same as Fig.~\ref{fig:Fixed_positionConc_NR} but for the reactive case at $D_a=1$,
see Fig.~\ref{fig:reactive_case_Da1_Pe100_1000}. 
}}
\label{fig:Fixed_positionConc}
\end{center}
\end{figure}

\subsection{Quantitative analysis}
\label{subsec:quantitative}

\begin{figure}[!htbp]
\begin{center}
\includegraphics[scale=0.36]{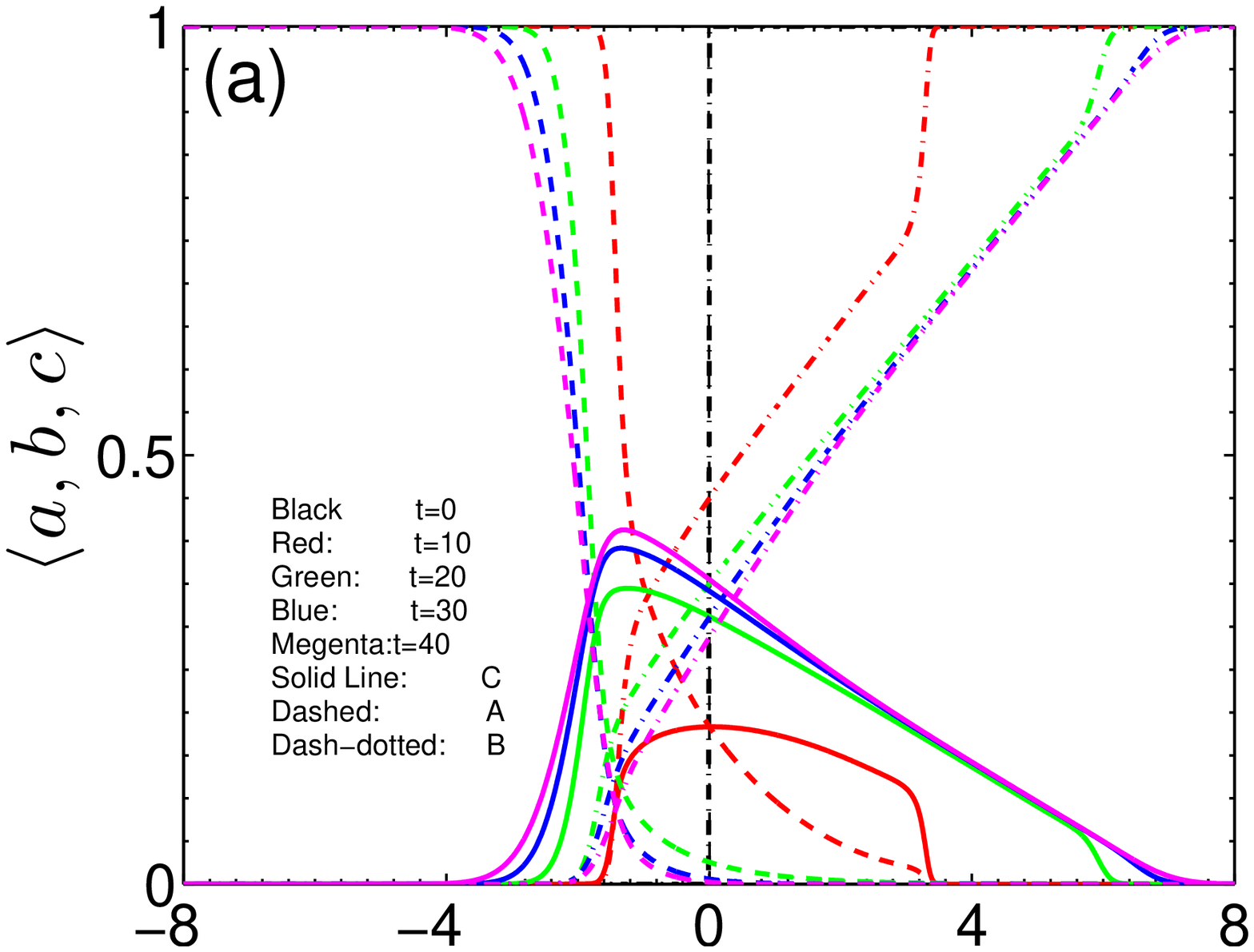}
\includegraphics[scale=0.35]{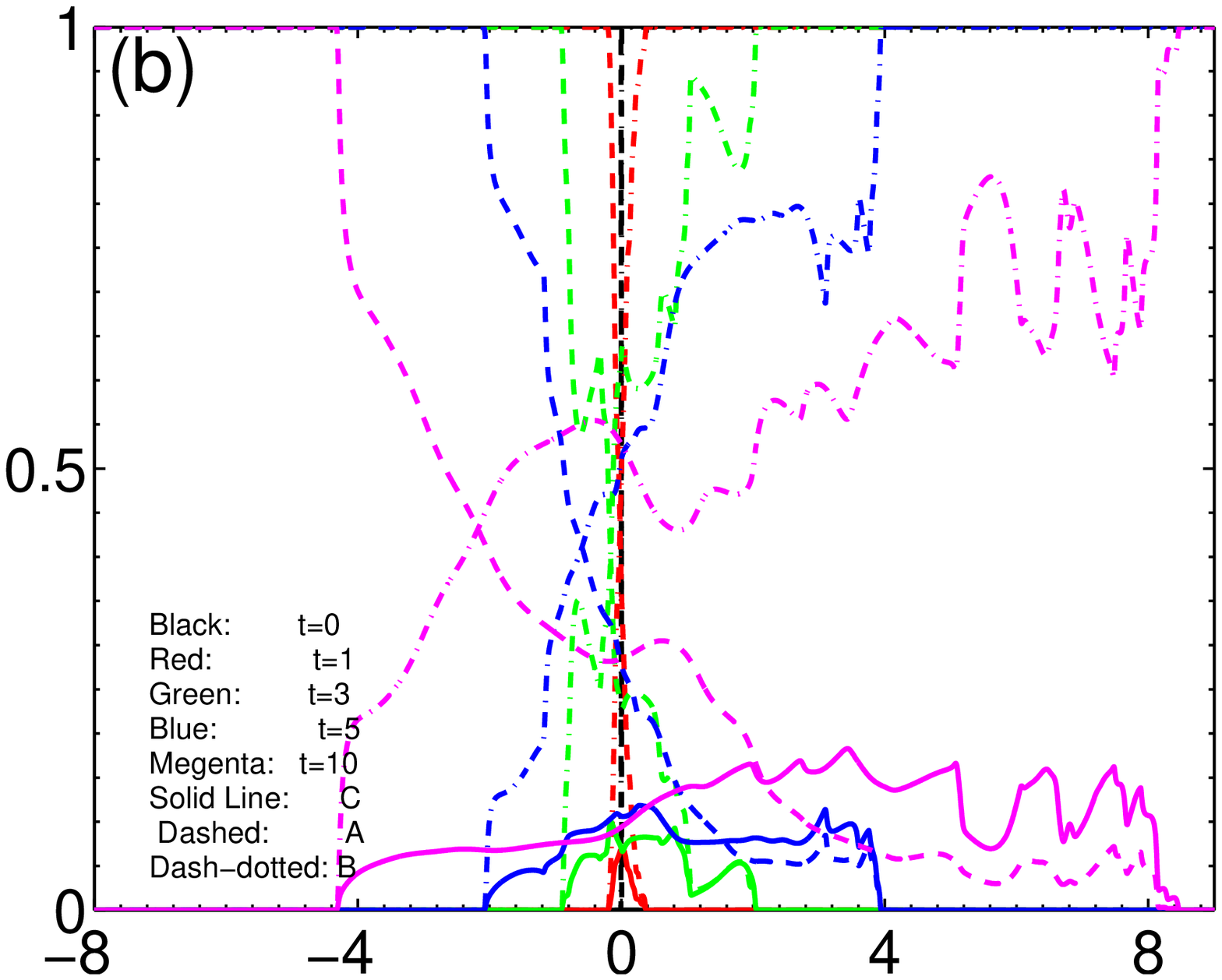}
 \includegraphics[scale=0.35]{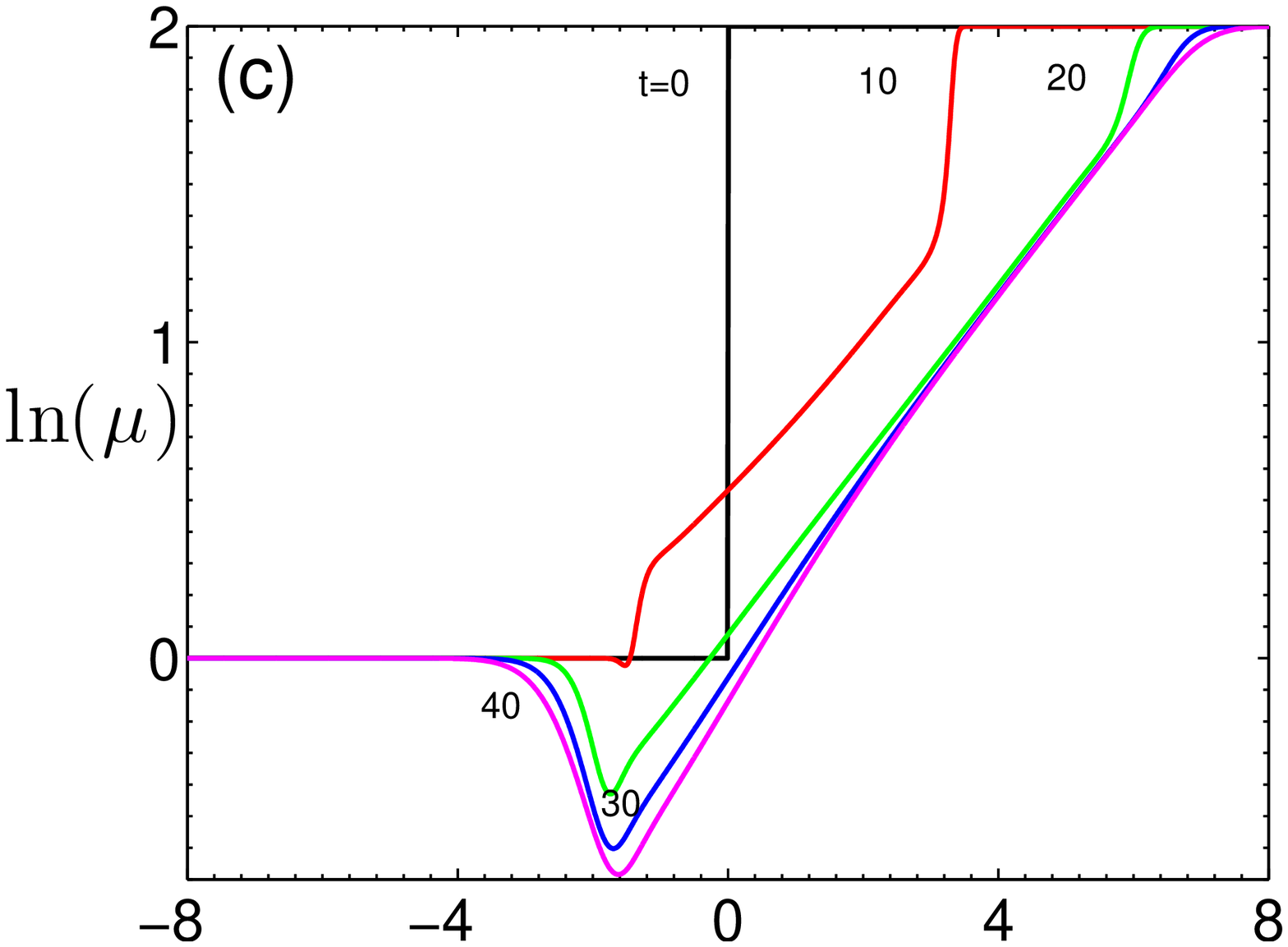}\quad
 \includegraphics[scale=0.35]{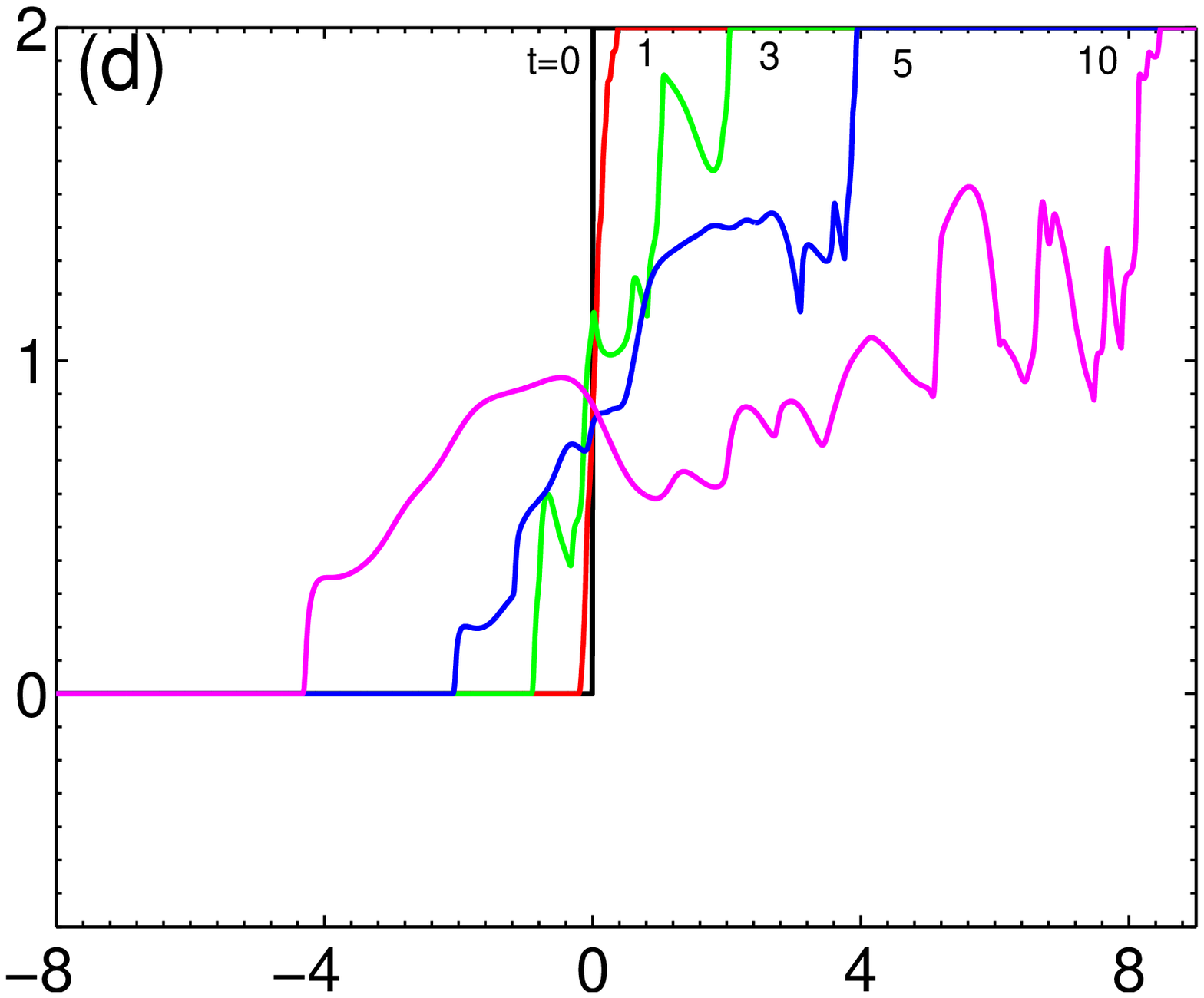} 
  \includegraphics[scale=0.35]{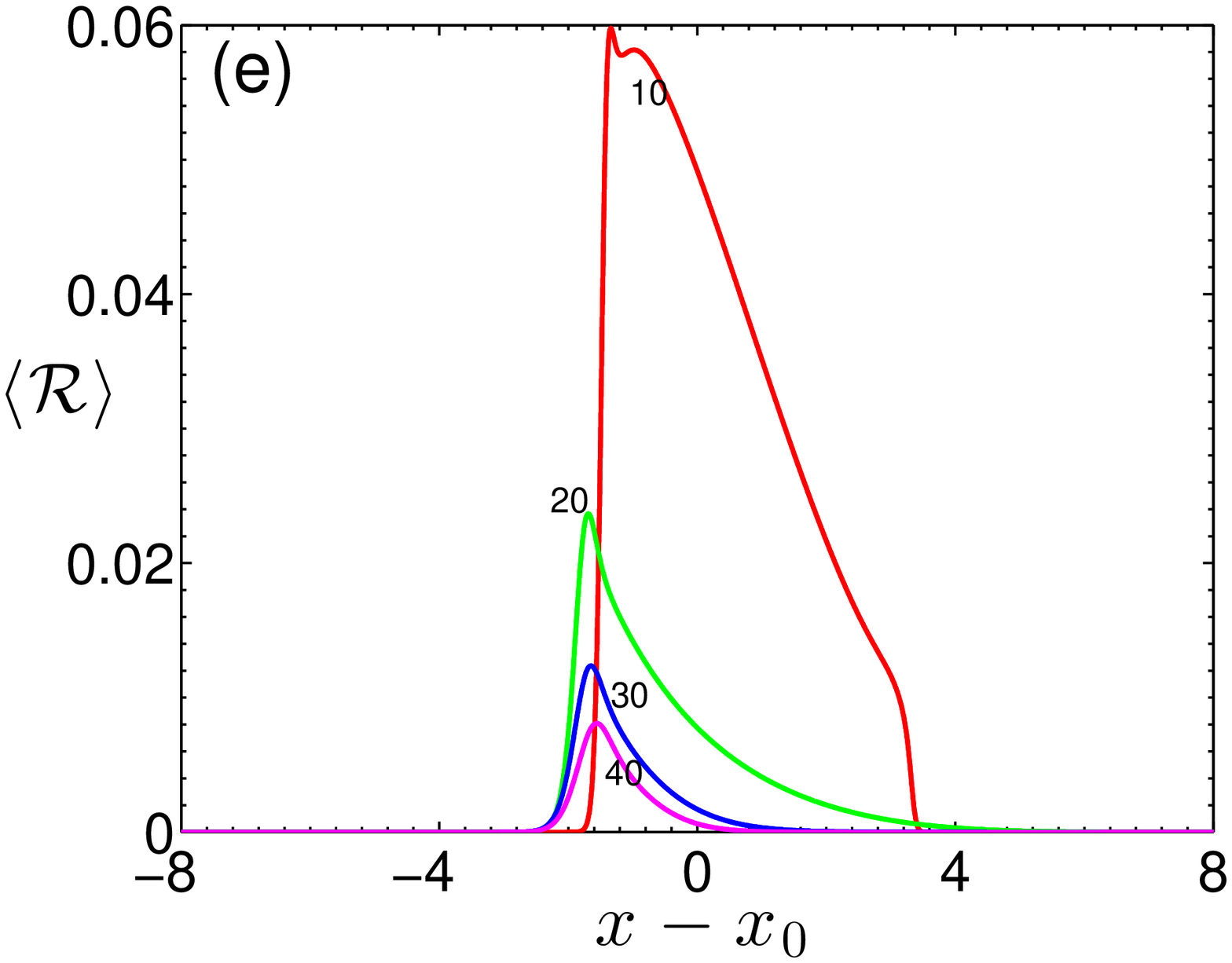}\quad
 \includegraphics[scale=0.35]{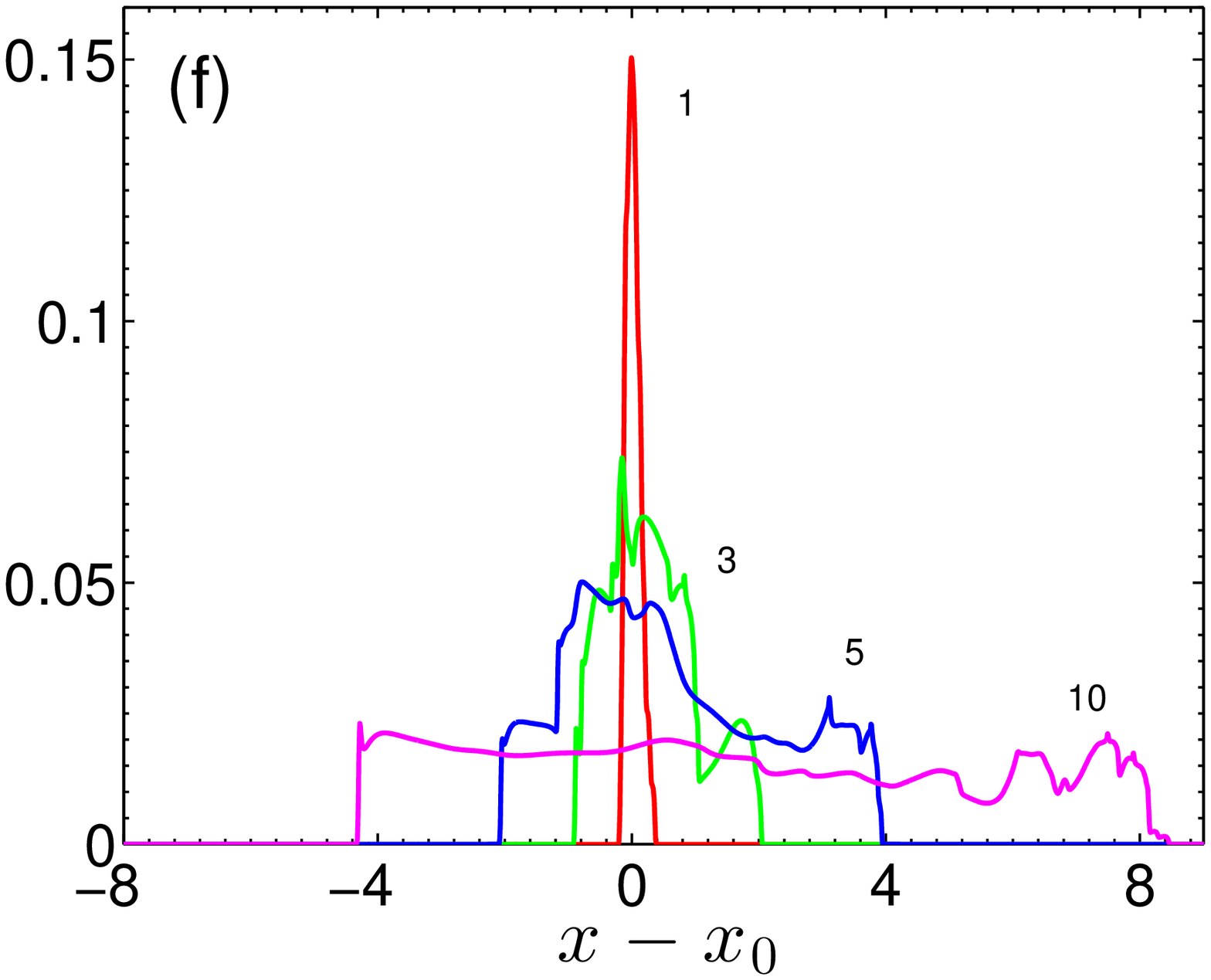}
\caption{\small{
Transversely averaged concentration (top row), viscosity  (middle row), and reaction rate (bottom row) profiles corresponding to simulations of Figs.~\ref{fig:reactive_case_Da1_Pe100_1000} for ${\rm Pe}=100$ (left column) and  ${\rm Pe}=1000$ (right column).
The dashed, dash-dotted and solid lines in panels (a,b) depict concentrations of $A$, $B$ and $C$, respectively.
The black, red, green, blue and magenta colors in 
panels~(a,c,e) correspond to $t=0$, $10$, $20$, $40$ and $50$, respectively, while those in panels (b,d,f) correspond to $t=0$, $1$, $3$, $5$ and $10$, respectively. 
}}
\label{fig:profile_Pe100Pe1000_Da1}
\end{center}
\end{figure}

In order to understand the opposite dynamics at low and high Pe, and to quantify the influence of varying ${\rm Pe}$ on reactive VF, we compute the  one-dimensional transversely averaged profiles of given quantities, $\zeta(x,y,t)$ as
\begin{equation}
\langle \zeta(x,t) \rangle= \frac{1}{L_y} \int_{0}^{L_y}  \zeta(x,y,t)\,\,{\rm d}y,
\label{eqn:profile}
\end{equation}
where $\zeta$ can be, for instance, concentration, viscosity, etc. In absence of fingering ($R_b=R_c=0$), these profiles are equivalent to the one-dimensional reaction diffusion profiles. For the simulations of Fig.~\ref{fig:reactive_case_Da1_Pe100_1000}, the temporal evolution of some of these transversely averaged profiles  is shown in  Fig.~\ref{fig:profile_Pe100Pe1000_Da1}. 

In the convective flow regime, the fingering pattern starts to develop around the reactive interface as soon as solutions $A$ and $B$ react and produce a less-viscous product $C$, see Figs.~\ref{fig:profile_Pe100Pe1000_Da1}(a) and \ref{fig:profile_Pe100Pe1000_Da1}(b). As the system evolves in time, we see that increasing amounts of  $A$ and $B$ are consumed and that the total production of the product $\langle c(x,t) \rangle$ increases. The corresponding reaction rate $\langle \mathcal{R}(x,t) \rangle$, shown in Fig.~\ref{fig:profile_Pe100Pe1000_Da1}(e), decreases in time when $A$ and $B$ are consumed and are progressively separated by $C$. Fig.~\ref{fig:profile_Pe100Pe1000_Da1}(c) shows the development of viscosity as time evolves. At low ${\rm Pe}$, a viscosity minimum develops in time at the back of the reaction front where the product concentration is maximum which can also be seen from Figs.~\ref{fig:reactive_case_Da1_Pe100_1000}(c) and~\ref{fig:reactive_case_Da1_Pe100_1000}(d). Owing to the viscosity minimum, the interface between $A$ and $C$ is stabilized, which can clearly be observed in Fig.~\ref{fig:reactive_case_Da1_Pe100_1000}(a) as the interface tends to flatten. On the contrary, the interface between $B$ and $C$ where the less viscous $C$ pushes the more viscous $B$  indicates the presence of VF. Nevertheless, transverse diffusion finally dominates VF, and  the interface between $B$ and $C$ eventually stabilizes again  [see Fig.~\ref{fig:reactive_case_Da1_Pe100_1000}(a--e)].

Let us now analyze quantitatively fingering patterns at larger ${\rm Pe}$. We have noticed in Fig.~\ref{fig:reactive_case_Da1_Pe100_1000}(f--j) that reactive VF is destabilizing at high ${\rm Pe}$ in contrast to a stabilizing trend at low ${\rm Pe}$. Figures~\ref{fig:profile_Pe100Pe1000_Da1}(b,d,f) show that, at high ${\rm Pe}$, when VF is present, the transversely averaged concentration profiles feature bumps indicating the presence of forward and backward fingering. In contrast to fingering at the back,  forward fingering shows merging and tip-splitting, see Fig.~\ref{fig:reactive_case_Da1_Pe100_1000}(f--j). Similar to the concentration, the log-viscosity, Fig.~\ref{fig:profile_Pe100Pe1000_Da1}(d), and reaction rate profiles, Fig.~\ref{fig:profile_Pe100Pe1000_Da1}(f), show similar features. The center of mass of these profiles is shifted towards the right of the reaction front indicating the presence of more elongated fingering in the $B$-rich region. While, at low ${\rm Pe}$, the viscosity minimum formed at the back (or left) of the reaction front gives rise to stabilization, it is completely absent at high ${\rm Pe}$ causing VF to expand significantly around the reaction zone.

\section{Parametric Study}
\label{sec:parametric_analysis}

We have seen that fingering is  stabilized at lower ${\rm Pe}$ when the viscosity decreases thanks to a chemical reaction.  To gain more insight into this stabilization effect, a parametric study is next carried out at  several low ${\rm Pe}$ values to  understand the effect of varying the Damk{\" o}hler number $D_a$ and the viscosity of the product by changing the log-mobility ratio $R_c$.    

\subsection{Effect of mobility ratio $R_c$ at $R_b>0$}
\label{subsec:effRc}

\begin{figure}[!htbp]
\begin{center}
 \includegraphics[scale=0.33]{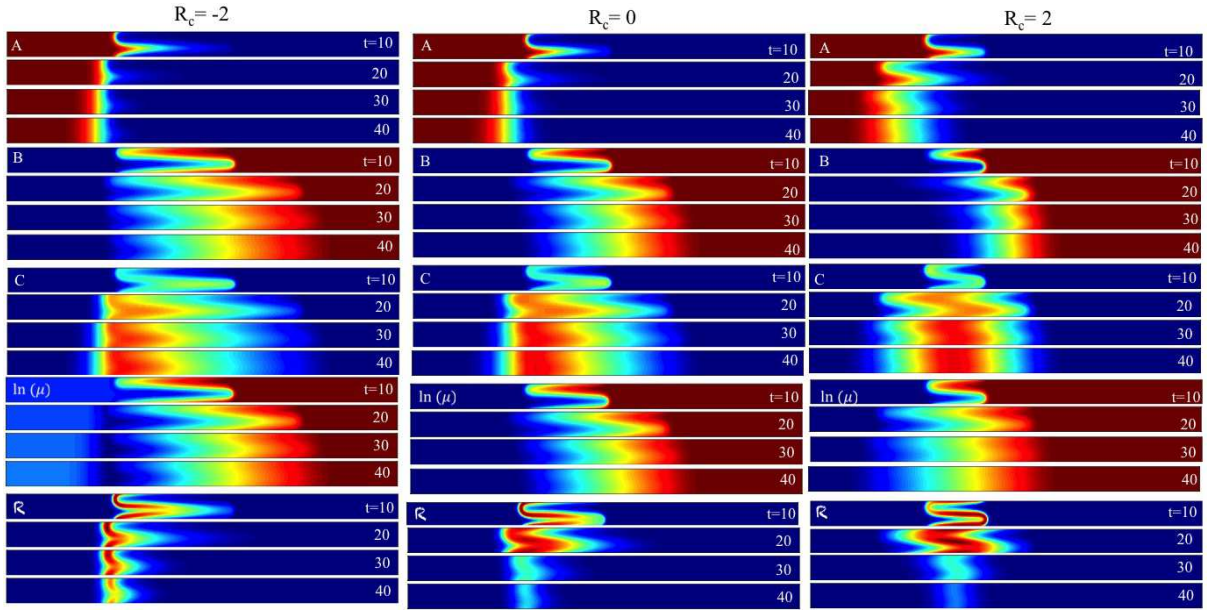}
\caption{\small{From top to bottom in each column: concentrations of $A$, $B$ and $C$, $\mbox{ln}(\mu)$ and $\mathcal{R}$ for various time steps ($t=10$, $20$, $30$ and $40$) at ${\rm Pe}=100$ with $R_c=-2$ (first column),  $R_c=0$ (second column) and $R_c=2$ (equivalent of the non-reactive case, third column). Other parameter values are as in Fig.~\ref{fig:reactive_case_Da1_Pe100_1000}.
}}
\label{fig:EffR}
\end{center}
\end{figure}

The effect of changing the log-mobility ratio $R_c$ is shown on Fig.~\ref{fig:EffR}. We consider the three values $R_c=-2$, $0$, $2$. We remind that, when $R_b=R_c$ (=2 here), the consumption of $B$ is balanced by the production of $C$, hence the dynamics of the reactive case is  equivalent to that of the non reactive system. When $0<R_C<R_b$, the viscosity decreases by the reaction but the viscosity profile remains monotonic in space. On the contrary, if $R_c<0$, a minimum in viscosity develops in time. For $R_b=2$, the cases  $R_c=2$, $R_c=0$ and $R_c=-2$ represent thus the (i) non-reactive VF, (ii) reactive VF with monotonic viscosity, and (iii) reactive VF with a viscosity minimum, respectively. 

By comparing concentrations of $A$, $B$ and $C$ for various time steps in these three cases, we see that, when $R_c<0$, the viscosity minimum has the following effects: (i) The interface between $A$ and $C$ stabilizes rapidly
and the mixing of reactant $A$ decreases as compared to the other two cases, (ii) as time evolves, the mixing region between $C$ and $B$ increases and stops fingering in time, displacing more $B$ by the product $C$. The reactive VF is  stabilized at low ${\rm Pe}$ by the  viscosity minimum compared to the  reactive VF case with monotonic viscosity or the  non-reactive VF.

\begin{figure}[!htbp]
\begin{center}
   \includegraphics[scale=0.5]{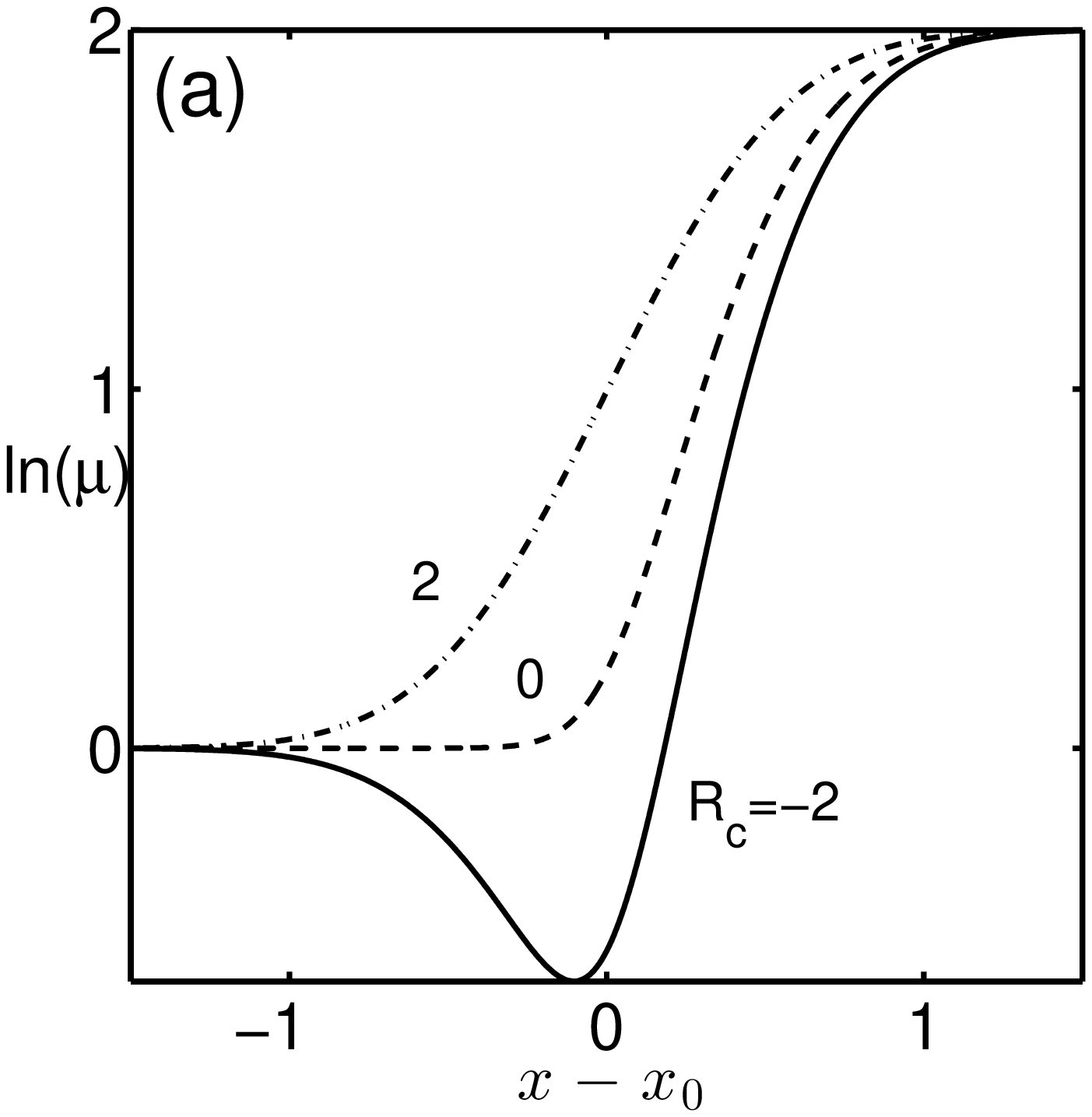}\qquad
   \includegraphics[scale=0.5]{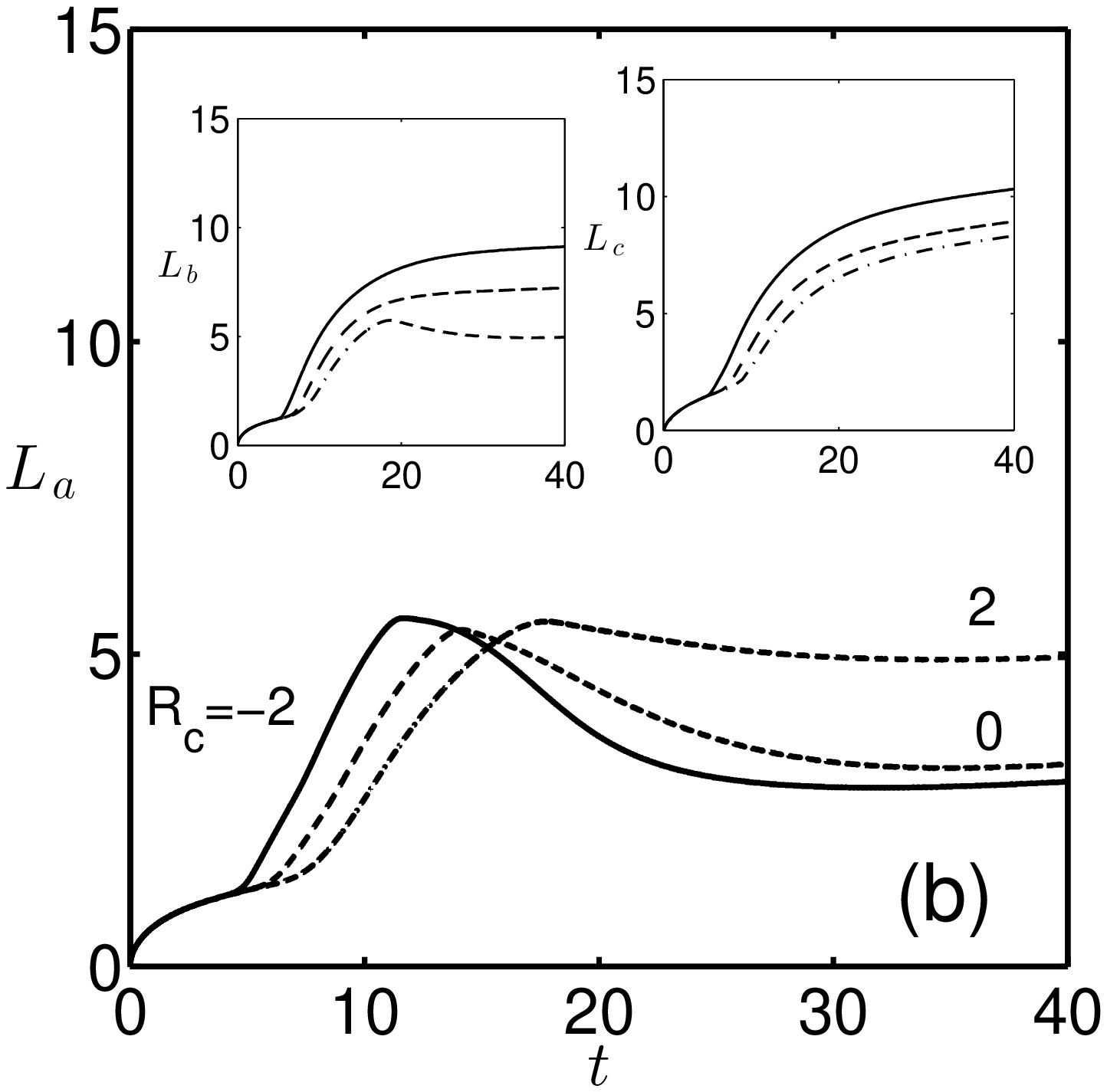}.
  \caption{\small{(a) RD profiles for $\mbox{ln}(\mu)$, and (b) temporal variation of the mixing length for 
 $A$ (main panel), $B$ (inset), and $C$ (inset) for $R_c:$ $-2$ (solid line), $0$ (dashed line), and $2$ (dash-dotted line). Other parameter values are as in Fig.~\ref{fig:EffR}.
 }}
\label{fig:effR_2_RD_L}
\end{center}
\end{figure}

The origin of this stabilization can be explained through the long time asymptotic one-dimensional reaction diffusion (RD) profiles of ${\rm ln}(\mu)$, as shown in Fig.~\ref{fig:effR_2_RD_L}(a). If $R_c=-2$, the reaction diffusion viscosity front moves in time from the higher viscosity region of $B$ to the lower viscosity region of $A$, see Fig.~\ref{fig:profile_Pe100Pe1000_Da1}(a). Due to the presence of lower viscosity region containing $C$, the profile of ${\rm ln}(\mu)$ develops a minimum in the $A$-rich region. While the gradients ${\rm d}({\rm ln}\mu)/{\rm d}x$ are decreasing with $R_c$ on the left of the reaction front [$x-x_0<0$], those  on the right [$x-x_0>0$] are increasing. Owing to this, when $R_c=-2$, the miscible interface between $A$ and $C$ is more stable as is the case when a higher viscosity fluid displaces a lower viscosity one. As a consequence, the mixing length $L_a$ decreases rapidly as time evolves and finally reaches a steady value which is the lowest among all  cases, as shown in the main panel of Fig.~\ref{fig:effR_2_RD_L}(b).

In contrast to the interface between $A$ and $C$, the interface between $C$ and $B$ is more unstable when $R_c=-2$ because ${\rm d}({\rm ln}\mu)/{\rm d}x$ is then the steepest, see Fig.~\ref{fig:effR_2_RD_L}(a). This can also be noticed in the evolution of the mixing length of $B$ and $C$ in Fig.~\ref{fig:effR_2_RD_L}(b). The instability at the interface between $B$ and $C$ starts earlier and the mixing length $L_b$ and $L_c$ increase more in time as $R_c$ decreases. As time evolves (far from the onset), due to  transverse diffusion, $L_b$ and $L_c$ reach a steady value which increases with decreasing $R_c$. The displacement of $B$ is thus larger when $R_c=-2$ in comparison to $R_c=0$ and $2$ (non-reactive). From Figs.~\ref{fig:reactive_case_Da1_Pe100_1000}--\ref{fig:effR_2_RD_L}, we can thus conclude that, when $R_c=-2$, the front between $A$ and $C$ stabilizes, the mixing between $B$ and $C$ is increased and the displacement of $B$ is larger.

\subsection{Effect of P{\'e}clet number $Pe$}
\label{subsec:effPe}

\begin{figure}[!htbp]
\begin{center}
\includegraphics[scale=0.3]{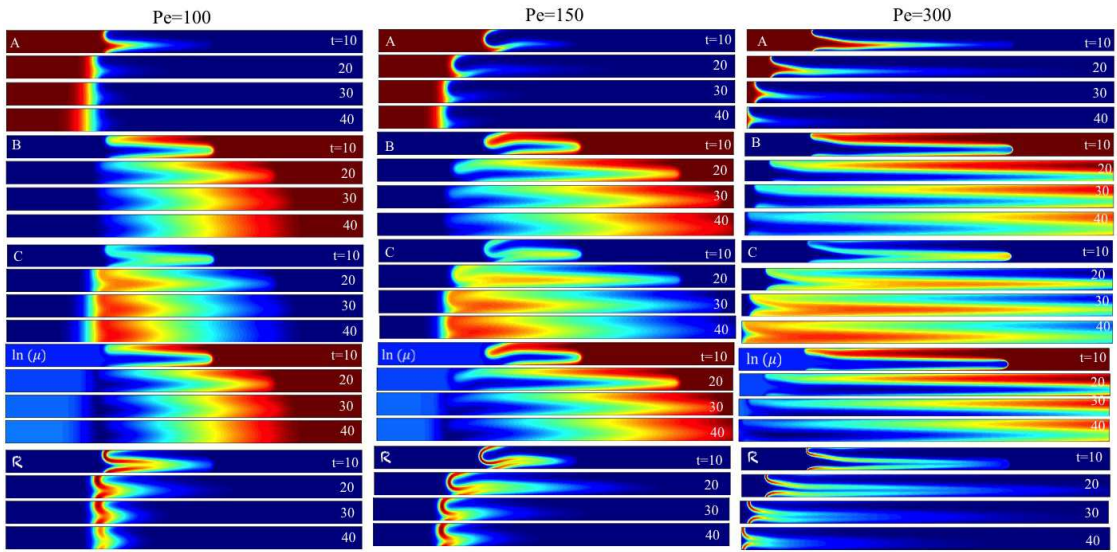}
\caption{\small{Same as Fig.~\ref{fig:EffR} but for $R_c=-2$ and  different values of ${\rm Pe}$: $100$ (first column), $150$ (second column) and $300$ (third column).}}
\label{fig:effPe}
\end{center}
\end{figure}

\begin{figure}[!htbp]
\begin{center}
    \includegraphics[scale=0.45]{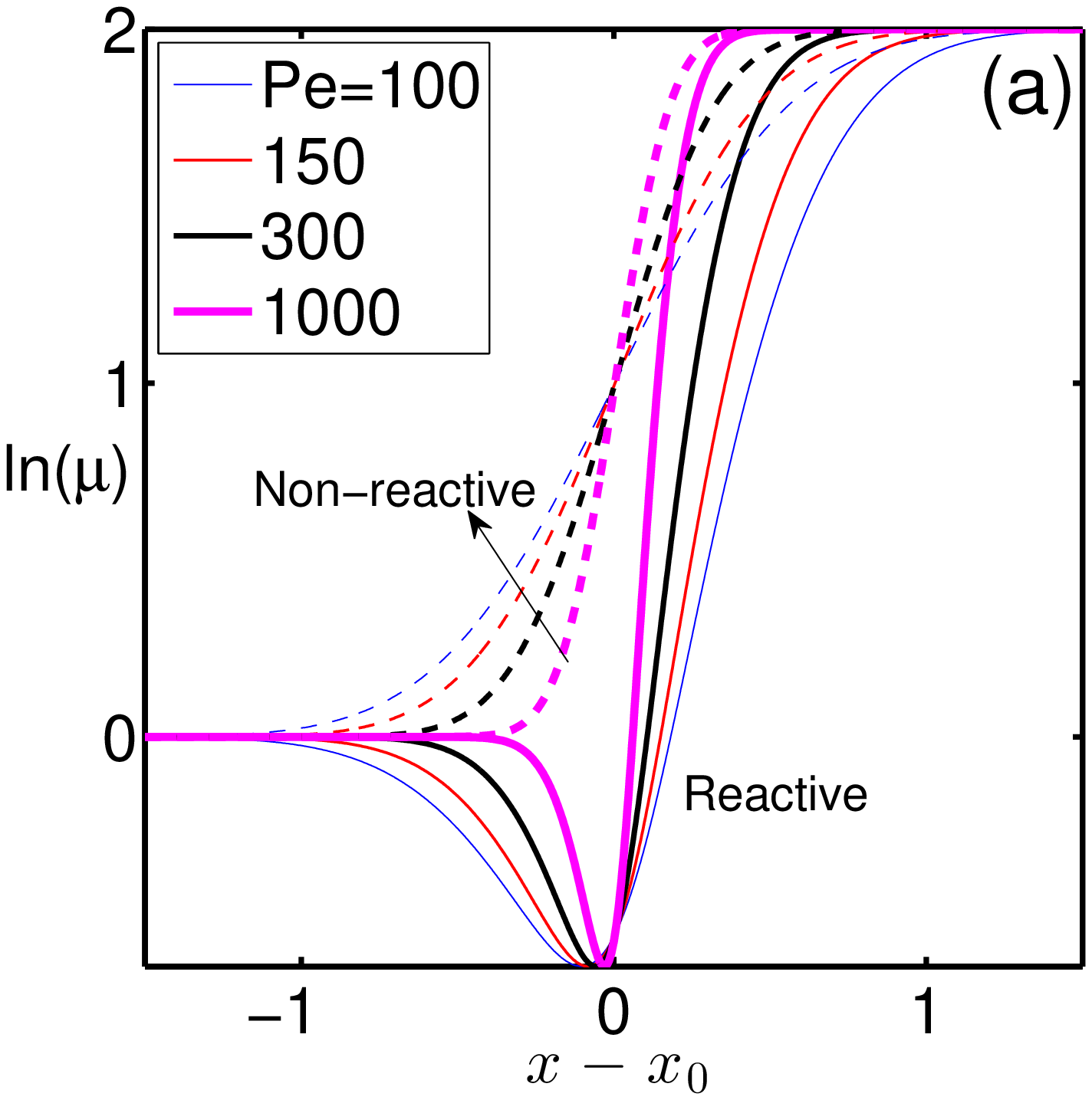}\quad
     \includegraphics[scale=0.45]{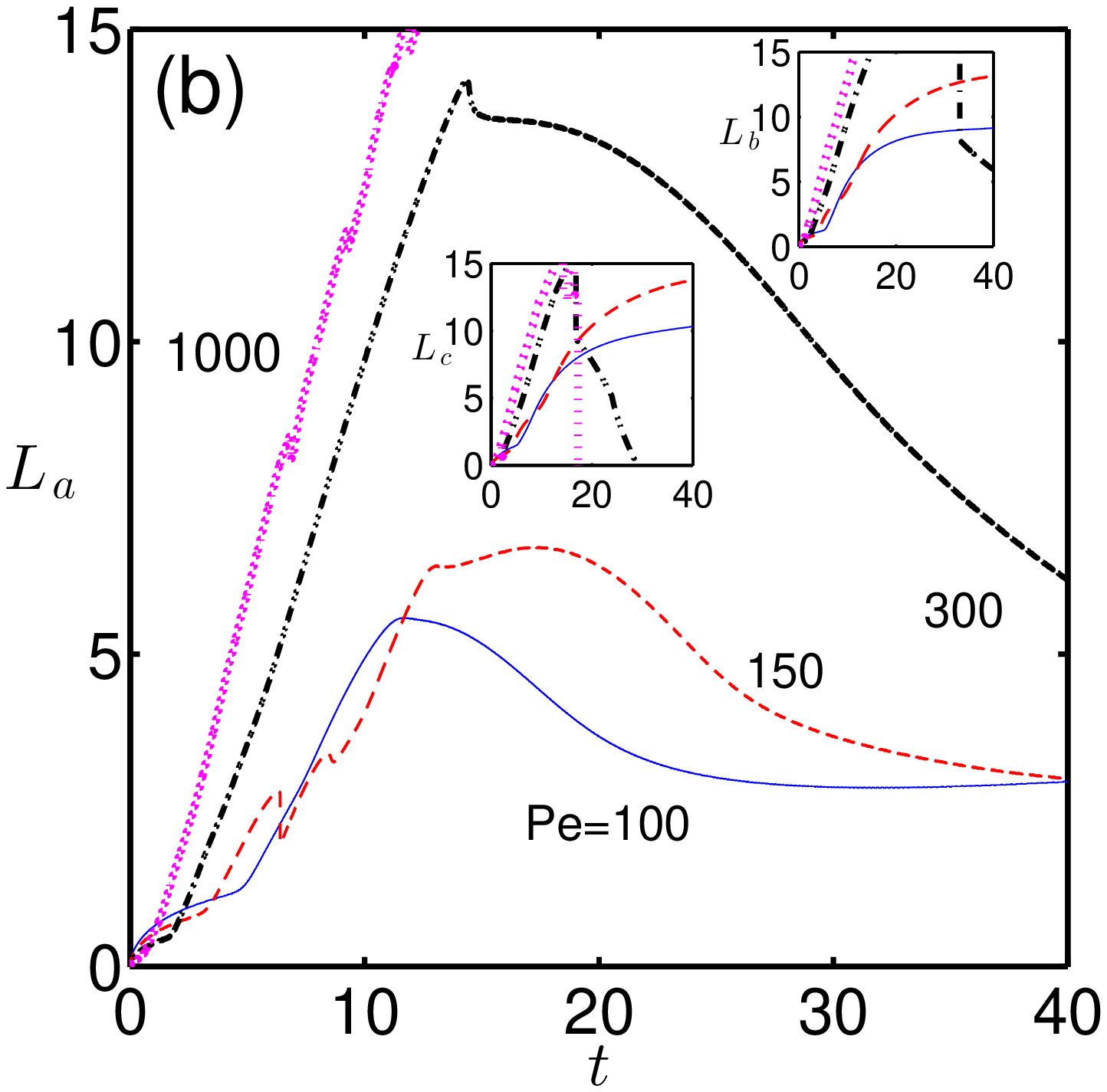}
 \caption{\small{(a) RD profiles of $\mbox{ln}(\mu)$ for 
 ${\rm Pe}=100$ (blue),~$150$ (red),~$300$ (black) and $1000$ (magenta) where the solid and dashed lines represent reactive and non-reactive systems, respectively. (b) Temporal evolution of the mixing length for $A$ (main panel), $B$ (upper inset) and $C$ (lower inset) at ${\rm Pe}=100$ (blue solid line), $150$ (red dashed line), $300$ (black dash-dotted line) and $1000$ (magenta dotted line). Other parameter values are as in Fig.~\ref{fig:effPe}.}}
\label{fig:effPe_2}
\end{center}
\end{figure}

In the previous section, we have seen that the onset time decreases i.e. the system is initially more unstable as $R_c$ decreases. We now fix $R_c=-2$ and analyze the effect on fingering of changing ${\rm Pe}$ keeping it nevertheless at small values. Specifically, concentrations, $\mbox{ln}(\mu)$ and the reaction rate are shown for ${\rm Pe}:$~$100$ (first column), $150$ (second column) and $300$ (third column) in Fig.~\ref{fig:effPe}. We see that the system becomes more unstable when increasing ${\rm Pe}$. This can be understood by inspecting the one-dimensional RD profiles shown in Fig.~\ref{fig:effPe_2}(a). The non-reactive displacement (dashed lines) is more unstable at higher ${\rm Pe}$ because the gradient of viscosity ${\rm d}({\rm ln}\mu)/{\rm d}x$ is correspondingly sharper. Similarly, viscosity gradients in the reactive RD systems (solid lines) are larger when ${\rm Pe}$ increases as diffusion is then less efficient to smooth the viscosity profile. Consequently, the RDC system also becomes more unstable with increasing ${\rm Pe}$, as shown in Fig.~\ref{fig:effPe} and on the evolution of the mixing lengths, see Fig.~\ref{fig:effPe_2}(b) where we see that the onset time of the fingering instability decreases with increasing ${\rm Pe}$. The smaller ${\rm Pe}$, the quicker the mixing lengths tend to a steady state value at low ${\rm Pe}$ whereas at large ${\rm Pe}$ the mixing lengths are increasing instantaneously irrespective of the viscosity minimum at the interface. 

\subsection{Effect of Damk{\"o}hler number, $Da$}
\label{subsec:effDa}

\begin{figure}[!h]
\begin{center}
\includegraphics[scale=0.33]{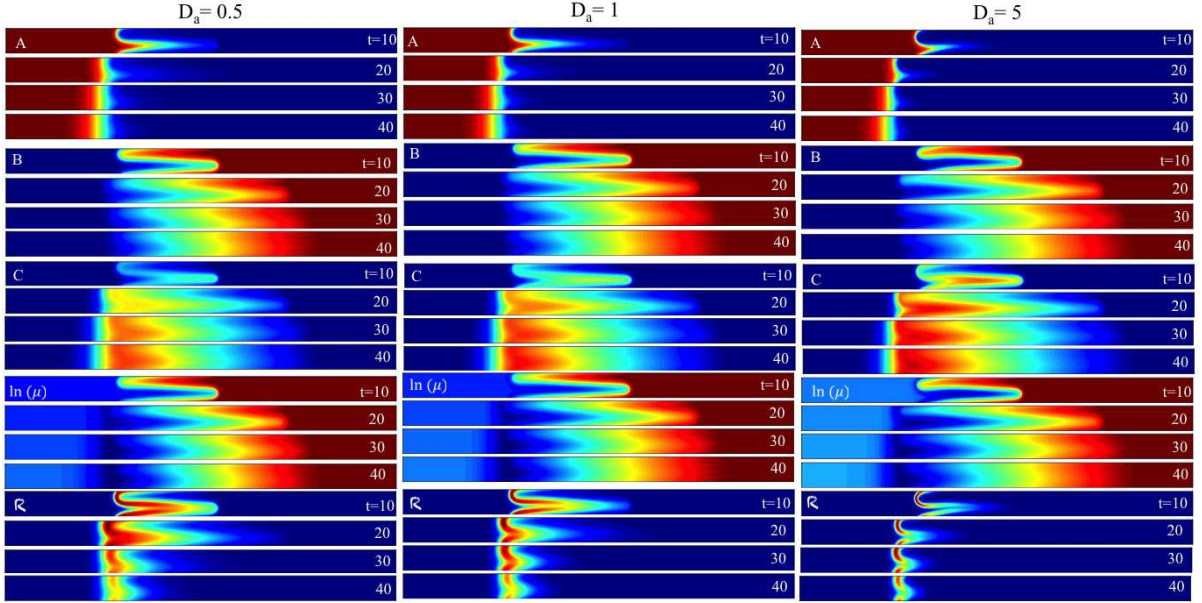}
 \caption{\small{
 Same as Fig.~\ref{fig:EffR} but for ${\rm Pe}=100$ and different values of ${\rm D_a}$:
$0.5$ (first column), $1$ (second column) and $5$ (third column). 
 }}
\label{fig:effDa}
\end{center}
\end{figure}
\begin{figure}[!htbp]
\begin{center}
 \includegraphics[scale=0.45]{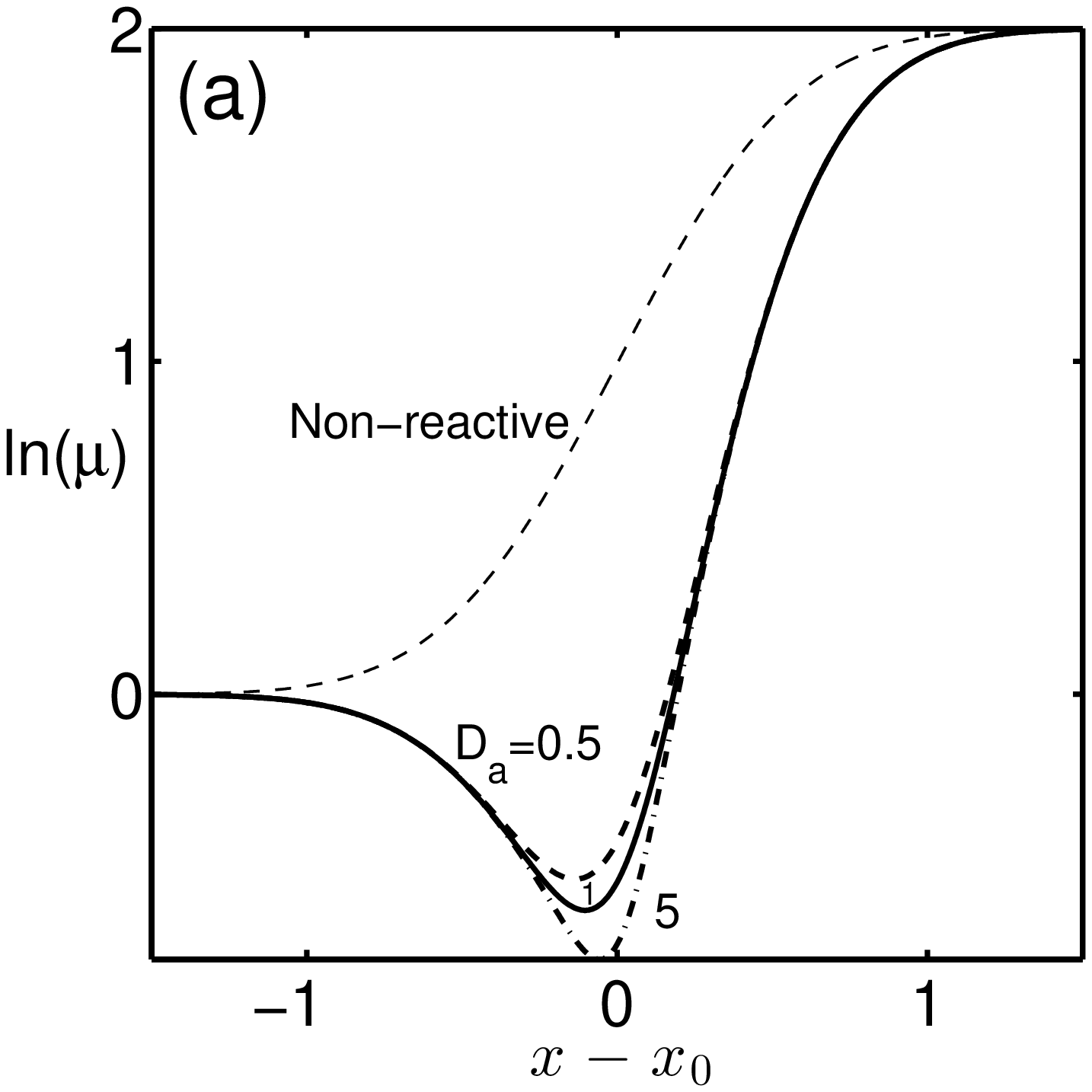}\qquad
 \includegraphics[scale=0.45]{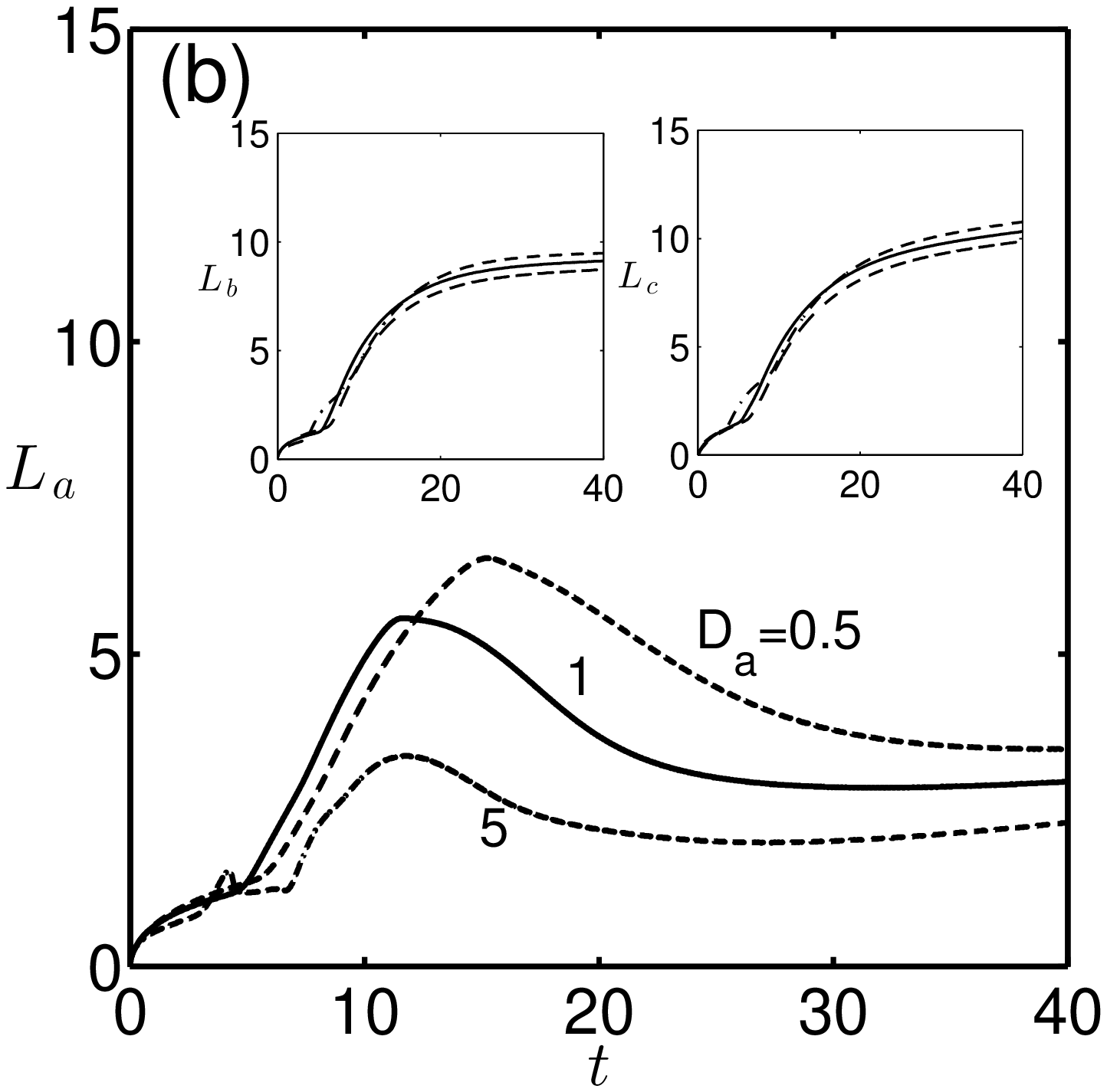}
 \caption{\small{Same as Fig.~\ref{fig:effPe_2} for variable $D_a$: $0.5$ (thick dashed line), $1.0$ (thick solid line) and $5.0$ (thick dot-dashed line). The thin dashed line in panel (a) represents the non-reactive case.  
}}
\label{fig:effDa_2}
\end{center}
\end{figure}

To study the effect of varying the Damk{\"o}hler number on the stabilization of fingering instability thanks to reactions  decreasing the viscosity, Figure~\ref{fig:effDa} depicts the concentrations, ${\rm ln}(\mu)$ and the reaction rate $\mathcal{R}$ at successive times for three values of $D_a$. We see that, when increasing $D_a$ (i.e.~the reaction occurs faster), the viscosity minimum develops more quickly (see also Fig.~\ref{fig:effDa_2}(a)), the amount of product $C$ formed at a given time increases, and the  reaction rate $\mathcal{R}$ decays faster because the reactants $A$ and $B$ are increasingly separated by the product $C$. As a consequence, when $D_a$ increases, the miscible interface between $A$ and $C$ stabilizes faster and the steady value of $L_a$ decreases. In parallel, the interface between $B$ and $C$ becomes  uniform in time, and the corresponding values of $L_b$ and $L_c$ saturates (see Fig.~\ref{fig:effDa_2}(b)). The system is thus globally more stable when $D_a$ is larger. 

We conclude thus from this parametric study that the displacement tends to stabilize (destabilize) at lower ${\rm Pe}$ (high ${\rm Pe}$) for $R_c<0$ ($R_c \geq 0$), and larger $D_a$ (smaller $D_a$). The optimal conditions to avoid fingering can thus be achieved when the viscosity is decreasing by a fast chemical reaction provided the rate of injection of displacing fluid is kept as low as possible to allow the viscosity minimum to build up.

\section{Conclusion and Outlook}
\label{sec:Conclusion and Outlook}

We have here analysed the influence of the injection flow rate on reactive VF driven by a simple $A+B \rightarrow C$ type chemical reaction decreasing the viscosity {\it in situ}. To do so, we have numerically integrated Darcy's law for the evolution of the flow velocity and RDC equations for the concentrations coupled by a viscosity profile depending dynamically on the concentration of the  chemical species. The injection flow rate has been varied by changing the values of the dimensionless parameter ${\rm Pe}$. Nonlinear simulations have been performed to characterise  the properties of reactive VF when a solution of a reactant $A$ displaces a solution of $B$ to produce the less viscous product $C$ at the miscible reactive interface. At lower ${\rm Pe}$, the VF instability is less intense in both reactive and non reactive cases because the viscosity gradients are smoothed out by diffusion. The reactive VF pattern covers nevertheless a larger area i.e. is spatially denser than the non-reactive pattern. These observations are in good agreement with experiments  \cite{riolfo,NMKT2007}. Similarly to the non-reactive case, at higher ${\rm Pe}$, VF is enhanced in reactive systems when the viscosity minimum does not have time to build up. Less-dense fingering patterns and more mixing are then observed. In other words, the fingering patterns at high ${\rm Pe}$ cover a smaller area than at low ${\rm Pe}$. In terms of  displacement efficiency, the presence of a viscosity minimum at lower ${\rm Pe}$ is found to optimize a homogeneous and regular displacement with less convective mixing. 

Our study provides a mathematical framework to control VF in many geophysical processes e.g.~reactive pollutant displacement, ${\rm CO}_2$ sequestration and EOR. Recently, it has been shown that fingering instabilities in the application of EOR can be controlled by introducing a viscosity minimum in the zone of contact between the two fluids via  the formation of foam between the injected gas and displaced oil \cite{Farajzadeh2015}. In this context, the present study (i) provides a convection between viscosity minimum and stabilization, (ii) introduces a way to control VF by controlling the injection rate, (iii) shows that, at low injection rate, the reactive VF improves the sweep efficiency in comparison to the non reactive conditions.

\section*{Acknowledgment}
We thank Y. Nagatsu, F. Brau, F. Haudin and M. Mishra for fruitful discussions. 
P.S. acknowledges financial support from IIT Madras for a New Faculty Initiation Grant, and a New Faculty Seed Grant.
A.D. acknowledges PRODEX for financial support.

\bibliography{refer}

\begin{thebibliography}{36}
\expandafter\ifx\csname natexlab\endcsname\relax\def\natexlab#1{#1}\fi
\expandafter\ifx\csname bibnamefont\endcsname\relax
  \def\bibnamefont#1{#1}\fi
\expandafter\ifx\csname bibfnamefont\endcsname\relax
  \def\bibfnamefont#1{#1}\fi
\expandafter\ifx\csname citenamefont\endcsname\relax
  \def\citenamefont#1{#1}\fi
\expandafter\ifx\csname url\endcsname\relax
  \def\url#1{\texttt{#1}}\fi
\expandafter\ifx\csname urlprefix\endcsname\relax\def\urlprefix{URL }\fi
\providecommand{\bibinfo}[2]{#2}
\providecommand{\eprint}[2][]{\url{#2}}

\bibitem[{\citenamefont{Saffman and Taylor}(1958)}]{ST1958}
\bibinfo{author}{\bibfnamefont{P.~G.} \bibnamefont{Saffman}} \bibnamefont{and}
  \bibinfo{author}{\bibfnamefont{G.~I.} \bibnamefont{Taylor}}, ``The
  penetration of a fluid into a porous medium or {Hele-Shaw} cell containing a
  more viscous liquid,'' \bibinfo{journal}{Proc. R. Soc. Lond. A}
  \textbf{\bibinfo{volume}{245}}, \bibinfo{pages}{312} (\bibinfo{year}{1958}).

\bibitem[{\citenamefont{Saffman}(1986)}]{S1986}
\bibinfo{author}{\bibfnamefont{P.~G.} \bibnamefont{Saffman}}, ``The penetration
  of a fluid into a porous medium or {Hele-Shaw} cell containing a more viscous
  liquid,'' \bibinfo{journal}{J. Fluid Mech.} \textbf{\bibinfo{volume}{173}},
  \bibinfo{pages}{73} (\bibinfo{year}{1986}).

\bibitem[{\citenamefont{Tan and Homsy}(1986)}]{TH1986}
\bibinfo{author}{\bibfnamefont{C.~T.} \bibnamefont{Tan}} \bibnamefont{and}
  \bibinfo{author}{\bibfnamefont{G.~M.} \bibnamefont{Homsy}}, ``Stability of
  miscible displacements in porous media: Rectilinear flow,''
  \bibinfo{journal}{Phys. Fluids} \textbf{\bibinfo{volume}{29}},
  \bibinfo{pages}{3549} (\bibinfo{year}{1986}).

\bibitem[{\citenamefont{Homsy}(1987)}]{Homsy1987}
\bibinfo{author}{\bibfnamefont{G.~M.} \bibnamefont{Homsy}}, ``Viscous fingering
  in porous media,'' \bibinfo{journal}{Annu. Rev. Fluid Mech.}
  \textbf{\bibinfo{volume}{19}}, \bibinfo{pages}{271} (\bibinfo{year}{1987}).

\bibitem[{\citenamefont{Tan and Homsy}(1988)}]{TH1988}
\bibinfo{author}{\bibfnamefont{C.~T.} \bibnamefont{Tan}} \bibnamefont{and}
  \bibinfo{author}{\bibfnamefont{G.~M.} \bibnamefont{Homsy}}, ``Simulation of
  nonlinear viscous fingering in miscible displacement,''
  \bibinfo{journal}{Phys. Fluids} \textbf{\bibinfo{volume}{31}},
  \bibinfo{pages}{1330} (\bibinfo{year}{1988}).

\bibitem[{\citenamefont{Orr and Taber}(1984)}]{OT1984}
\bibinfo{author}{\bibfnamefont{F.~M.} \bibnamefont{Orr}} \bibnamefont{and}
  \bibinfo{author}{\bibfnamefont{J.~J.} \bibnamefont{Taber}}, ``Use of carbon
  dioxide in enhanced oil recovery,'' \bibinfo{journal}{Science}
  \textbf{\bibinfo{volume}{224}}, \bibinfo{pages}{563} (\bibinfo{year}{1984}).

\bibitem[{\citenamefont{Lunn and Kueper}(1999)}]{LK1999}
\bibinfo{author}{\bibfnamefont{S.~R.} \bibnamefont{Lunn}} \bibnamefont{and}
  \bibinfo{author}{\bibfnamefont{B.~H.} \bibnamefont{Kueper}}, ``Manipulation
  of density and viscosity for the optimization of {DNAPL} recovery by alcohol
  flooding,'' \bibinfo{journal}{J. Contam. Hydrol.}
  \textbf{\bibinfo{volume}{38}}, \bibinfo{pages}{427 } (\bibinfo{year}{1999}).

\bibitem[{\citenamefont{Fu et~al.}(2013)\citenamefont{Fu, Cueto-Felgueroso, and
  Juanes}}]{FCJ2013}
\bibinfo{author}{\bibfnamefont{X.}~\bibnamefont{Fu}},
  \bibinfo{author}{\bibfnamefont{L.}~\bibnamefont{Cueto-Felgueroso}},
  \bibnamefont{and} \bibinfo{author}{\bibfnamefont{R.}~\bibnamefont{Juanes}},
  ``Pattern formation and coarsening dynamics in three-dimensional convective
  mixing in porous media,'' \bibinfo{journal}{Phil. Trans. R. Soc.}
  \textbf{\bibinfo{volume}{371}} (\bibinfo{year}{2013}).

\bibitem[{\citenamefont{Farajzadeh et~al.}(2015)\citenamefont{Farajzadeh,
  Eftekhari, Hajibeygi, van~der Meer, Vincent-Bonnieu, and
  Rossen}}]{Farajzadeh2015}
\bibinfo{author}{\bibfnamefont{R.}~\bibnamefont{Farajzadeh}},
  \bibinfo{author}{\bibfnamefont{A.}~\bibnamefont{Eftekhari}},
  \bibinfo{author}{\bibfnamefont{H.}~\bibnamefont{Hajibeygi}},
  \bibinfo{author}{\bibfnamefont{J.}~\bibnamefont{van~der Meer}},
  \bibinfo{author}{\bibfnamefont{S.}~\bibnamefont{Vincent-Bonnieu}},
  \bibnamefont{and} \bibinfo{author}{\bibfnamefont{W.}~\bibnamefont{Rossen}},
  in \emph{\bibinfo{booktitle}{SPE Reservoir Simulation Symposium, Houston,
  Texas, USA, 23–25 February 2015}} (\bibinfo{publisher}{Society of Petroleum
  Engineers}, \bibinfo{address}{Texas, USA}, \bibinfo{year}{2015}), pp.
  \bibinfo{pages}{SPE--173193--MS}.

\bibitem[{\citenamefont{Ritsema et~al.}(1998)\citenamefont{Ritsema, Dekker,
  Nieber, and Steenhuis}}]{RDNS1998}
\bibinfo{author}{\bibfnamefont{C.~J.} \bibnamefont{Ritsema}},
  \bibinfo{author}{\bibfnamefont{L.~W.} \bibnamefont{Dekker}},
  \bibinfo{author}{\bibfnamefont{J.~L.} \bibnamefont{Nieber}},
  \bibnamefont{and} \bibinfo{author}{\bibfnamefont{T.~S.}
  \bibnamefont{Steenhuis}}, ``Modeling and field evidence of finger formation
  and finger recurrence in a water repellent sandy soil,''
  \bibinfo{journal}{Water Resour. Res.} \textbf{\bibinfo{volume}{34}},
  \bibinfo{pages}{555} (\bibinfo{year}{1998}).

\bibitem[{\citenamefont{Hornof and Baig}(1995)}]{HB1995}
\bibinfo{author}{\bibfnamefont{V.}~\bibnamefont{Hornof}} \bibnamefont{and}
  \bibinfo{author}{\bibfnamefont{F.~U.} \bibnamefont{Baig}}, ``Influence of
  interfacial reaction and mobility ratio on the displacement in a hele-shaw
  cell,'' \bibinfo{journal}{Exp. Fluids} \textbf{\bibinfo{volume}{18}},
  \bibinfo{pages}{448} (\bibinfo{year}{1995}).

\bibitem[{\citenamefont{Jahodab and Hornofa}(2000)}]{JH2000}
\bibinfo{author}{\bibfnamefont{M.}~\bibnamefont{Jahodab}} \bibnamefont{and}
  \bibinfo{author}{\bibfnamefont{V.}~\bibnamefont{Hornofa}}, ``Concentration
  profiles of reactant in a viscous finger formed during the interfacially
  reactive immiscible displacements in porous media,'' \bibinfo{journal}{Powder
  Technology} \textbf{\bibinfo{volume}{110}}, \bibinfo{pages}{253}
  (\bibinfo{year}{2000}).

\bibitem[{\citenamefont{Nagatsu and Ueda}(2001)}]{NU2001}
\bibinfo{author}{\bibfnamefont{Y.}~\bibnamefont{Nagatsu}} \bibnamefont{and}
  \bibinfo{author}{\bibfnamefont{T.}~\bibnamefont{Ueda}}, ``Effects of reactant
  concentrations on reactive miscible viscous fingering,''
  \bibinfo{journal}{AIChE J.} \textbf{\bibinfo{volume}{47}},
  \bibinfo{pages}{1711} (\bibinfo{year}{2001}).

\bibitem[{\citenamefont{Nagatsu and Ueda}(2003)}]{NU2003}
\bibinfo{author}{\bibfnamefont{Y.}~\bibnamefont{Nagatsu}} \bibnamefont{and}
  \bibinfo{author}{\bibfnamefont{T.}~\bibnamefont{Ueda}}, ``Effects of
  finger-growth velocity on reactive miscible viscous fingering,''
  \bibinfo{journal}{AIChE J.} \textbf{\bibinfo{volume}{49}},
  \bibinfo{pages}{789} (\bibinfo{year}{2003}).

\bibitem[{\citenamefont{Nagatsu
  et~al.}(2009{\natexlab{a}})\citenamefont{Nagatsu, Ogawa, Kato, and
  Tada}}]{NOKT2009}
\bibinfo{author}{\bibfnamefont{Y.}~\bibnamefont{Nagatsu}},
  \bibinfo{author}{\bibfnamefont{T.}~\bibnamefont{Ogawa}},
  \bibinfo{author}{\bibfnamefont{Y.}~\bibnamefont{Kato}}, \bibnamefont{and}
  \bibinfo{author}{\bibfnamefont{Y.}~\bibnamefont{Tada}}, ``Investigation of
  reacting flow fields in miscible viscous fingering by a novel experimental
  method,'' \bibinfo{journal}{AIChE J.} \textbf{\bibinfo{volume}{55}},
  \bibinfo{pages}{563} (\bibinfo{year}{2009}{\natexlab{a}}).

\bibitem[{\citenamefont{De~Wit and Homsy}(1999{\natexlab{a}})}]{WH1999}
\bibinfo{author}{\bibfnamefont{A.}~\bibnamefont{De~Wit}} \bibnamefont{and}
  \bibinfo{author}{\bibfnamefont{G.~M.} \bibnamefont{Homsy}}, ``Viscous
  fingering in reaction-diffusion systems,'' \bibinfo{journal}{J. Chem. Phys.}
  \textbf{\bibinfo{volume}{110}}, \bibinfo{pages}{8663}
  (\bibinfo{year}{1999}{\natexlab{a}}).

\bibitem[{\citenamefont{De~Wit and Homsy}(1999{\natexlab{b}})}]{WH1999a}
\bibinfo{author}{\bibfnamefont{A.}~\bibnamefont{De~Wit}} \bibnamefont{and}
  \bibinfo{author}{\bibfnamefont{G.~M.} \bibnamefont{Homsy}}, ``Nonlinear
  interactions of chemical reactions and viscous fingering in porous media,''
  \bibinfo{journal}{Phys. Fluids} \textbf{\bibinfo{volume}{11}},
  \bibinfo{pages}{949} (\bibinfo{year}{1999}{\natexlab{b}}).

\bibitem[{\citenamefont{Fernandez and Homsy}(2003)}]{FH2003}
\bibinfo{author}{\bibfnamefont{J.}~\bibnamefont{Fernandez}} \bibnamefont{and}
  \bibinfo{author}{\bibfnamefont{G.~M.} \bibnamefont{Homsy}}, ``Viscous
  fingering with chemical reaction: effect of in-situ production of
  surfactants,'' \bibinfo{journal}{J. Fluid Mech.}
  \textbf{\bibinfo{volume}{480}}, \bibinfo{pages}{267} (\bibinfo{year}{2003}).

\bibitem[{\citenamefont{Podgorski et~al.}(2007)\citenamefont{Podgorski,
  Sostarecz, Zorman, and Belmonte}}]{PSZB2007}
\bibinfo{author}{\bibfnamefont{T.}~\bibnamefont{Podgorski}},
  \bibinfo{author}{\bibfnamefont{M.~C.} \bibnamefont{Sostarecz}},
  \bibinfo{author}{\bibfnamefont{S.}~\bibnamefont{Zorman}}, \bibnamefont{and}
  \bibinfo{author}{\bibfnamefont{A.}~\bibnamefont{Belmonte}}, ``Fingering
  instabilities of a reactive micellar interface,'' \bibinfo{journal}{Phys.
  Rev. E} \textbf{\bibinfo{volume}{76}}, \bibinfo{pages}{016202}
  (\bibinfo{year}{2007}).

\bibitem[{\citenamefont{Nagatsu et~al.}(2007)\citenamefont{Nagatsu, Matsuda,
  Kato, and Tada}}]{NMKT2007}
\bibinfo{author}{\bibfnamefont{Y.}~\bibnamefont{Nagatsu}},
  \bibinfo{author}{\bibfnamefont{K.}~\bibnamefont{Matsuda}},
  \bibinfo{author}{\bibfnamefont{Y.}~\bibnamefont{Kato}}, \bibnamefont{and}
  \bibinfo{author}{\bibfnamefont{Y.}~\bibnamefont{Tada}}, ``Experimental study
  on miscible viscous fingering involving viscosity changes induced by
  variations in chemical species concentrations due to chemical reactions,''
  \bibinfo{journal}{J. Fluid Mech.} \textbf{\bibinfo{volume}{571}},
  \bibinfo{pages}{475} (\bibinfo{year}{2007}).

\bibitem[{\citenamefont{G\'erard and De~Wit}(2009)}]{GW2009}
\bibinfo{author}{\bibfnamefont{T.}~\bibnamefont{G\'erard}} \bibnamefont{and}
  \bibinfo{author}{\bibfnamefont{A.}~\bibnamefont{De~Wit}}, ``Miscible viscous
  fingering induced by a simple { $A+B \rightarrow C$ } chemical reaction,''
  \bibinfo{journal}{Phys. Rev. E} \textbf{\bibinfo{volume}{79}},
  \bibinfo{pages}{016308} (\bibinfo{year}{2009}).

\bibitem[{\citenamefont{Nagatsu
  et~al.}(2009{\natexlab{b}})\citenamefont{Nagatsu, Kondo, Kato, and
  Tada}}]{NKKT2009}
\bibinfo{author}{\bibfnamefont{Y.}~\bibnamefont{Nagatsu}},
  \bibinfo{author}{\bibfnamefont{Y.}~\bibnamefont{Kondo}},
  \bibinfo{author}{\bibfnamefont{Y.}~\bibnamefont{Kato}}, \bibnamefont{and}
  \bibinfo{author}{\bibfnamefont{Y.}~\bibnamefont{Tada}}, ``Effects of moderate
  damk{\"o}hler number on miscible viscous fingering involving viscosity
  decrease due to a chemical reaction,'' \bibinfo{journal}{J. Fluid Mech}
  \textbf{\bibinfo{volume}{625}}, \bibinfo{pages}{97}
  (\bibinfo{year}{2009}{\natexlab{b}}).

\bibitem[{\citenamefont{Hejazi et~al.}(2010)\citenamefont{Hejazi, Trevelyan,
  Azaiez, and De~Wit}}]{HTAW2010}
\bibinfo{author}{\bibfnamefont{S.~H.} \bibnamefont{Hejazi}},
  \bibinfo{author}{\bibfnamefont{P.~M.~J.} \bibnamefont{Trevelyan}},
  \bibinfo{author}{\bibfnamefont{J.}~\bibnamefont{Azaiez}}, \bibnamefont{and}
  \bibinfo{author}{\bibfnamefont{A.}~\bibnamefont{De~Wit}}, ``Viscous fingering
  of a miscible reactive {$A+B \rightarrow C$} interface: a linear stability
  analysis,'' \bibinfo{journal}{J. Fluid Mech.} \textbf{\bibinfo{volume}{652}},
  \bibinfo{pages}{501} (\bibinfo{year}{2010}).

\bibitem[{\citenamefont{Hejazi and Azaiez}(2010{\natexlab{a}})}]{HA2010a}
\bibinfo{author}{\bibfnamefont{S.~H.} \bibnamefont{Hejazi}} \bibnamefont{and}
  \bibinfo{author}{\bibfnamefont{J.}~\bibnamefont{Azaiez}}, ``Non-linear
  interactions of dynamic interfaces in porous media,'' \bibinfo{journal}{Chem.
  Eng. Sci.} \textbf{\bibinfo{volume}{65}}, \bibinfo{pages}{938}
  (\bibinfo{year}{2010}{\natexlab{a}}).

\bibitem[{\citenamefont{Hejazi and Azaiez}(2010{\natexlab{b}})}]{HA2010b}
\bibinfo{author}{\bibfnamefont{S.~H.} \bibnamefont{Hejazi}} \bibnamefont{and}
  \bibinfo{author}{\bibfnamefont{J.}~\bibnamefont{Azaiez}}, ``Hydrodynamic
  instability in the transport of miscible reactive slices through porous
  media,'' \bibinfo{journal}{Phys. Rev. E} \textbf{\bibinfo{volume}{81}},
  \bibinfo{pages}{056321} (\bibinfo{year}{2010}{\natexlab{b}}).

\bibitem[{\citenamefont{Nagatsu and De~Wit}(2011)}]{NW2011}
\bibinfo{author}{\bibfnamefont{Y.}~\bibnamefont{Nagatsu}} \bibnamefont{and}
  \bibinfo{author}{\bibfnamefont{A.}~\bibnamefont{De~Wit}}, ``Viscous fingering
  of a miscible reactive { $A+B \rightarrow C$ } interface for an infinitely
  fast chemical reaction: Nonlinear simulations,'' \bibinfo{journal}{Phys.
  Fluids} \textbf{\bibinfo{volume}{23}}, \bibinfo{pages}{043103}
  (\bibinfo{year}{2011}).

\bibitem[{\citenamefont{Nagatsu et~al.}(2011)\citenamefont{Nagatsu, Kondo,
  Kato, and Tada}}]{NKKT2011}
\bibinfo{author}{\bibfnamefont{Y.}~\bibnamefont{Nagatsu}},
  \bibinfo{author}{\bibfnamefont{Y.}~\bibnamefont{Kondo}},
  \bibinfo{author}{\bibfnamefont{Y.}~\bibnamefont{Kato}}, \bibnamefont{and}
  \bibinfo{author}{\bibfnamefont{Y.}~\bibnamefont{Tada}}, ``Miscible viscous
  fingering involving viscosity increase by a chemical reaction with moderate
  damk{\"o}hler number,'' \bibinfo{journal}{Phys. Fluids}
  \textbf{\bibinfo{volume}{23}}, \bibinfo{pages}{014109}
  (\bibinfo{year}{2011}).

\bibitem[{\citenamefont{Riolfo et~al.}(2012)\citenamefont{Riolfo, Nagatsu,
  Iwata, Maes, Trevelyan, and De~Wit}}]{RNIMTWit2012}
\bibinfo{author}{\bibfnamefont{L.~A.} \bibnamefont{Riolfo}},
  \bibinfo{author}{\bibfnamefont{Y.}~\bibnamefont{Nagatsu}},
  \bibinfo{author}{\bibfnamefont{S.}~\bibnamefont{Iwata}},
  \bibinfo{author}{\bibfnamefont{R.}~\bibnamefont{Maes}},
  \bibinfo{author}{\bibfnamefont{P.~M.~J.} \bibnamefont{Trevelyan}},
  \bibnamefont{and} \bibinfo{author}{\bibfnamefont{A.}~\bibnamefont{De~Wit}},
  ``Experimental evidence of reaction-driven miscible viscous fingering,''
  \bibinfo{journal}{Phys. Rev. E} \textbf{\bibinfo{volume}{85}},
  \bibinfo{pages}{015304} (\bibinfo{year}{2012}).

\bibitem[{\citenamefont{Riolfo}(2013)}]{riolfo}
\bibinfo{author}{\bibfnamefont{L.}~\bibnamefont{Riolfo}}, Ph.D. thesis,
  \bibinfo{school}{Universit{\'e} libre de Bruxelles, Brussels}
  (\bibinfo{year}{2013}).

\bibitem[{\citenamefont{Alhumade and Azaiez}(2013)}]{alh13}
\bibinfo{author}{\bibfnamefont{H.}~\bibnamefont{Alhumade}} \bibnamefont{and}
  \bibinfo{author}{\bibfnamefont{J.}~\bibnamefont{Azaiez}}, ``Stability
  analysis of reversible reactive flow displacements in porous media,''
  \bibinfo{journal}{Chem. Eng. Sci} \textbf{\bibinfo{volume}{101}},
  \bibinfo{pages}{46} (\bibinfo{year}{2013}).

\bibitem[{\citenamefont{Nagatsu}(2010)}]{Nagatsu2010}
\bibinfo{author}{\bibfnamefont{Y.}~\bibnamefont{Nagatsu}}, ``Viscous fingering
  phenomena with chemical reactions,'' \bibinfo{journal}{Chem. Eng. Sci}
  \textbf{\bibinfo{volume}{65}}, \bibinfo{pages}{938} (\bibinfo{year}{2010}).

\bibitem[{\citenamefont{Wit}(2016)}]{dew16}
\bibinfo{author}{\bibfnamefont{A.~D.} \bibnamefont{Wit}}, ``Chemo-hydrodynamic
  patterns in porous media,'' \bibinfo{journal}{Phil. Trans. Roy. Soc. A}
  \textbf{\bibinfo{volume}{374}}, \bibinfo{pages}{20150419}
  (\bibinfo{year}{2016}).

\bibitem[{\citenamefont{Nagatsu and Masumo}(2015)}]{NMW2015}
\bibinfo{author}{\bibfnamefont{Y.}~\bibnamefont{Nagatsu}} \bibnamefont{and}
  \bibinfo{author}{\bibfnamefont{T.}~\bibnamefont{Masumo}}, ``personal
  communication,'' \textbf{\bibinfo{volume}{--}} (\bibinfo{year}{2015}).

\bibitem[{\citenamefont{Mishra et~al.}(2007)\citenamefont{Mishra, Martin, and
  De~Wit}}]{MMW2007}
\bibinfo{author}{\bibfnamefont{M.}~\bibnamefont{Mishra}},
  \bibinfo{author}{\bibfnamefont{M.}~\bibnamefont{Martin}}, \bibnamefont{and}
  \bibinfo{author}{\bibfnamefont{A.}~\bibnamefont{De~Wit}}, ``Miscible viscous
  fingering with linear adsorption on the porous matrix,''
  \bibinfo{journal}{Phys. Fluids} \textbf{\bibinfo{volume}{19}},
  \bibinfo{eid}{073101} (\bibinfo{year}{2007}).

\bibitem[{\citenamefont{Pramanik and Mishra}(2015)}]{PM2015}
\bibinfo{author}{\bibfnamefont{S.}~\bibnamefont{Pramanik}} \bibnamefont{and}
  \bibinfo{author}{\bibfnamefont{M.}~\bibnamefont{Mishra}}, ``Effect of
  pe{\'c}let number on miscible rectilinear displacement in a hele-shaw cell,''
  \bibinfo{journal}{Phys. Rev. E} \textbf{\bibinfo{volume}{91}},
  \bibinfo{pages}{033006} (\bibinfo{year}{2015}).

\bibitem[{\citenamefont{Fornberg}(1998)}]{Fornberg1998}
\bibinfo{author}{\bibfnamefont{B.}~\bibnamefont{Fornberg}},
  \emph{\bibinfo{title}{A Practical Guide to Pseudospectral Methods}},
  Cambridge Monographs on Applied and Computational Mathematics
  (\bibinfo{publisher}{Cambridge University Press}, \bibinfo{year}{1998}).

\end{thebibliography}

\end{document}